*Article*

# Turbulence and Rossby Wave Dynamics with Realizable Eddy Damped Markovian Anisotropic Closure

**Jorgen S. Frederiksen [1]* and Terence J. O'Kane [2]**

[1] CSIRO Environment, Aspendale, Melbourne, 3195, Australia
[2] CSIRO Environment, Hobart, 7004, Australia; terence.o'kane@csiro.au
* Correspondence: jorgen.frederiksen@csiro.au

**Abstract:** The theoretical basis for the Eddy Damped Markovian Anisotropic Closure (EDMAC) is formulated for two-dimensional anisotropic turbulence interacting with Rossby waves in the presence of advection by a large-scale mean flow. The EDMAC is as computationally efficient as the Eddy Damped Quasi Normal Markovian (EDQNM) closure but in contrast is realizable in the presence of transient waves. The EDMAC is arrived at through systematic simplification of a generalization of the non-Markovian Direct Interaction Approximation (DIA) closure that has its origin in renormalized perturbation theory. Markovian Anisotropic Closures (MACs) are obtained from the DIA by using three variants of the Fluctuation Dissipation Theorem (FDT) with the information in the time history integrals instead carried by Markovian differential equations for two relaxation functions. One of the MACs is simplified to the EDMAC with analytical relaxation functions and high numerical efficiency, like the EDQNM. Sufficient conditions for the EDMAC to be realizable in the presence of Rossby waves are determined. Examples of the numerical integration of the EDMAC compared with the EDQNM are presented for two-dimensional isotropic and anisotropic turbulence, at moderate Reynolds numbers, possibly interacting with Rossby waves and large-scale mean flow. The generalization of the EDMAC for the statistical dynamics of other physical systems, to higher dimension and higher order nonlinearity is considered.

## 1. Introduction

Orszag [1] proposed the Eddy Damped Quasi Normal Markovian (EDQNM) model as an efficient and realizable closure for three-dimensional (3D) isotropic turbulence. The first numerical study of the EDQNM was made by Leith [2] who formulated and solved it for two-dimensional (2D) isotropic turbulence. Herring [3] derived a statistical dynamical closure for 2D anisotropic turbulence that generalized the EDQNM to situations without transient waves. The EDQNM is computationally efficient because it is Markovian and consists of just the equations for the single-time two-point cumulant in spectral space with an analytical expression for the triad relaxation function. However, as we discuss and analyze in detail in this article, the EDQNM is not guaranteed to be realizable for anisotropic turbulence interacting with transient waves. In contrast, the Eddy Damped Markovian Anisotropic Closure (EDMAC), for which we formulate the theoretical basis, is realizable.

One can arrive at the EDQNM in various ways as discussed for example by Lesieur [4] with our preference being from a reduction of the more fundamentally based Eulerian Direct Interaction Approximation (DIA) of Kraichnan [5]. The DIA is a more complex non-Markovian closure that is founded on renormalized perturbation theory (reviews of closures are presented in Refs. [4,6–13]). These reviews present analyses of the relative performance of the various closure models. The form of the eddy damping in the EDQNM is chosen to be consistent with the forward $k^{-\frac{5}{3}}$ energy cascading inertial range of 3D isotropic turbulence and the forward $k^{-3}$ enstrophy cascading inertial range of 2D turbulence.

Starting from the DIA, the EDQNM may be obtained by using a Markovian form for the response function and using the *current-time FDT* (Fluctuation Dissipation Theorem)

$$C_{\mathbf{k}}(t,t') = R_{\mathbf{k}}(t,t')C_{\mathbf{k}}(t,t) \tag{1}$$

for $t \geq t'$ and $C_{\mathbf{k}}(t,t') = C_{-\mathbf{k}}(t',t)$ for $t' > t$. In Equation (1), $R_{\mathbf{k}}(t,t')$ is the response function, $C_{\mathbf{k}}(t,t')$ is the two-time spectral cumulant and $C_{\mathbf{k}}(t,t)$ the current-time single-time cumulant at wavenumber (or strictly wavevector) $\mathbf{k}$. In the EDQNM, the response function is also specified by an analytical form consistent with either the 2D or 3D inertial ranges. With these modifications the time-history integrals of the DIA can be performed analytically to determine a triad relaxation function that enters the cumulant equation (see Section 7). In the EDQNM the strength of eddy damping is specified by an empirical constant as distinct from the DIA which has no empirical parameters.

The EDQNM closure has the advantage of being very computationally efficient since it scales like $O(T)$ where $T$ is the length of the time integration. This contrasts with the DIA non-Markovian closure, that scales as $O(T^3)$, because the two-point cumulants and response functions satisfy integro-differential equations. Although, we should note that the efficiency of the DIA can be improved to scaling like $O(T^2)$. This can be done by using the three-point cumulant in periodic restarts of the DIA closure numerical model [14–17].

The DIA is closely related to the Self Consistent Field Theory closure (SCFT) of Herring [18,19] and the Local Energy-Transfer Theory closure (LET) of McComb [6,20,21]. The SCFT and LET closures are also non-Markovian and, despite being originally derived by different methods, can in fact also be obtained by modifications of the DIA equations. These three Eulerian non-Markovian closures all have the same single-time two-point closure equation. However, the SCFT and LET use the *prior-time FDT* [12] (Equation (1)) defined by

$$C_{\mathbf{k}}(t,t') = R_{\mathbf{k}}(t,t')C_{\mathbf{k}}(t',t') \tag{2}$$

for $t \geq t'$ where $C_{\mathbf{k}}(t',t')$ is the prior-time single-time cumulant. In the SCFT, the response function, identical to that for the DIA, is taken as fundamental and the two-time cumulant is obtained from the prior-time FDT. For the LET the two-time cumulant, identical to that for the DIA, is fundamental and the response function is derived from the FDT in Equation (2). In this study we focus on the derivation of Markovian closures from the DIA closure, but we could equally consider the SCFT or LET as the starting points.

The EDQNM closure has been widely used in many studies and applications as reviewed in Refs. [4,9–12,22]. As emphasized in the detailed study of Bowman et al. [23] the EDQNM may not be realizable in the presence of transient waves such as drift waves in plasmas or equivalently Rossby waves in geophysical flows. In some studies, the waves have been assumed to be slowly varying and the quasi-steady state form of the triad relaxation function used to avoid possible singular behaviour [22–29]. In other studies, modified forms of quasi-normal Markovian closures have instead been used [30,31].

The reason that the EDQNM may not necessarily be realizable with transient waves is because the oscillations of the propagating waves mean that the real part of the triad relaxation functions in the EDQNM may on occasions become negative. Bowman et al. [23] found that the same problem also occurs if the prior time FDT in Equation (2) is used. However, they showed that a realizable Markovian closure could be obtained by using a FDT that is essentially a half-way house between the current-time and prior-time FDTs. This FDT, that we call the *correlation FDT* is given by

$$C_{\mathbf{k}}(t,t') = [C_{\mathbf{k}}(t,t)]^{\frac{1}{2}} R_{\mathbf{k}}(t,t')[C_{\mathbf{k}}(t',t')]^{\frac{1}{2}} \tag{3}$$

for $t \geq t'$. Bowman et al. [23] called their closure the Realizable Markovian Closure (RMC). Unlike the EDQNM closure, for which the triad relaxation functions have analytical expressions, the RMC relaxation functions need to be evaluated by solving a system of auxiliary differential equations. Hence the RMC is less computationally efficient than the EDQNM. Hu et al. [32] used the RMC in studies of the Hasegawa–Wakatani two-field equation of plasma physics. Bowman and Krommes [33] further developed and applied realizable Markovian closures for homogeneous turbulence in the presence of waves. They considered the interaction of drift waves (essentially Rossby waves) and anisotropic turbulence with the single-level Charney–Hasegawa–Mima model. They also formulated a realizable test-field model (RTFM), a generalization of Kraichnan's [34] test-field model (TFM), for turbulence in the presence of transient waves.

Traditionally, closure theory, including Markovian closures, like the EDQNM [1,2], have focused on the forward energy cascade of 3D turbulence and the forward enstrophy cascade of 2D turbulence in decaying and steady state systems. The EDQNM damping is normally specified to be consistent with these forward inertial ranges. Two-dimensional turbulence also has an inverse energy cascade with a steady state inertial range of $k^{-\frac{5}{3}}$ [35]. With differential rotation on a $\beta$–plane, supporting Rossby waves, the turbulence becomes anisotropic, with preferential zonal motion. This was noted by Rhines [36] based on theory and direct numerical simulations (DNS) and by Holloway and Hendershott [37] in studies with a modified TFM. Their TFM did not cater for transient Rossby waves but required an approximate steady state triad relaxation function to guarantee realizability [33]. Rhines [36] proposed that under suitable conditions the fluid velocities follow an inverse $k^{-5}$ power law in steady state where $k$ is the wavenumber, the magnitude of the total wavevector. Chekhlov et al. [38] performed a set of DNS experiments of 2D turbulence on a $\beta$–plane forced by small scale forcing. They found support for the scaling of Rhines, for the zonal components with $k_x = 0$ which followed a $k_y^{-5}$ power law. However, the rest of the spectrum followed the $k^{-\frac{5}{3}}$ inverse energy cascade, emphasizing the anisotropy induced by the Rossby waves [22,24–29,37]. This is further discussed in the review by Galperin et al. [39]. Krommes and Parker [40] also discuss the application of homogeneous closures to circulations with zonal flow generation through inverse cascades, such as on the giant planets and their moons, and to plasma physics (see also the reviews in Refs. [41,42]). They note however, that real flows are generally inhomogeneous and "statistically inhomogeneous solutions with nontrivial mean fields can emerge from a statistically homogeneous state by spontaneous symmetry breaking". Indeed, Frederiksen [43] shows that just an infinitesimal topographic perturbation can cause such symmetry breaking with the mountain torque driving a mean westward zonal flow on the $\beta$–plane and solid body rotation on the sphere. This symmetry breaking may considerably modify the structures of the larger scale flows in the inverse cascades.

For the earth's atmosphere, with a Rossby radius of deformation around 1,000 km, the larger scales of turbulent 2D and quasigeostrophic flows are determined by heating, topography and baroclinic instability that override the inverse cascades [4]. It is 2D homogeneous flows, in typical atmospheric parameter ranges, with forward enstrophy cascades, and the analogous 3D homogeneous flows with forward energy cascades, that are of primary interest in our current study. For flows in these parameter ranges the large-scale flows can be specified to have realistic initial spectra or driven towards such states with the smaller scales evolving towards their statistically steady states. In this way studies with homogeneous closures can throw considerable light onto the essential issues of the more complicated inhomogeneous turbulence interactions.

In the last few years, Markovian closures have also been developed for inhomogeneous turbulence interacting with mean flows, Rossby waves, and topography by Frederiksen and O'Kane [12,44,45]. Three types of Markovian Inhomogeneous Closures (MICs) were developed [44] as well as three types of abridged MICs [45]. For the abridged MICs, the mean field trajectory in the time history integrals is replaced by the current-time mean field and results in slightly more efficient closures. To date, all the MIC variants have been developed as modifications of the inhomogeneous non-Markovian QDIA closure [46–48].

The QDIA was developed for 2D inhomogeneous turbulent flows interacting with mean fields and topography by Frederiksen [46]. It was extended to include Rossby waves (Frederiksen and O'Kane [47]) and to multi-level and multi-field models for classical and quantum field theories (Frederiksen [48,49]). It was numerically implemented by O'Kane and Frederiksen [17] and subsequently used in further studies of turbulence dynamics, interactions with topography and Rossby waves, predictability of transitions between strong zonal flows and blocking, in data assimilation and for determining subgrid-scale parameterizations. The literature is reviewed in Refs. [45,49].

The MICs are obtained from the QDIA closure by assuming one of the three FDTs. They can be combined as:

$$C_{\mathbf{k}}(t,t') = [C_{\mathbf{k}}(t,t)]^{1-X} R_{\mathbf{k}}(t,t')[C_{\mathbf{k}}(t',t')]^{X} \tag{4}$$

for $t \geq t'$ and $C_{\mathbf{k}}(t,t') = C_{-\mathbf{k}}(t',t)$ for $t' > t$. The current-time FDT is recovered when $X = 0$, the prior-time FDT has $X = 1$ and the correlation FDT has $X = \frac{1}{2}$.

In the numerical studies of Frederiksen and O'Kane [44,45] it was found that both the MICs [44] and abridged MICs [45] closely reproduced the statistics of large ensembles of DNS in simulations of inhomogeneous turbulence interacting with mean flows, Rossby waves and topography at low Reynolds numbers. While the MIC and abridged MIC models are more computationally efficient than the non-Markovian QDIA closure, the relaxation functions still need to be calculated by solving auxiliary differential equations as for the RMC of Bowman et al. [23]. A still more efficient inhomogeneous Markovian closure with analytical relaxation functions, the Eddy Damped Markovian Inhomogeneous Closure (EDMIC) was also developed by Frederiksen and O'Kane [45].

In a recent tribute to Jack Herring [12], and review of his major achievements in statistical dynamical fluid dynamics, it was noted that it is possible to generalize the EDQNM to a closure that is realizable in the presence of transient Rossby waves interacting with anisotropic turbulence. The resulting Eddy Damped Markovian Anisotropic Closure (EDMAC) involves a frequency renormalization of the eddy damping and by construction is as efficient as the EDQNM.

The major aims of this article are as follows:

1. To provide a theoretical motivation for the realizable EDMAC based on renormalized perturbation theory;
2. To generalize the EDMAC model for the interaction of anisotropic turbulence with transient Rossby waves in the presence of transient large-scale flows;
3. To establish conditions under which the real part of the triad relaxation functions is positive semi-definite when the large-scale flow and Doppler shifted Rossby wave frequency have general time dependencies;
4. To examine the extent to which the frequency-dependent contribution to the eddy damping in the EDMAC model changes the evolved energy and palinstrophy spectra and Reynolds number and skewness, compared with the EDQNM, for rapidly evolving moderate Reynolds number turbulence interacting with Rossby waves.

The methodology we apply for developing the realizable EDMAC model has parallels with the approach used by Frederiksen [46] to formulate the inhomogeneous quasi-diagonal direct interaction approximation (QDIA) closure. Frederiksen [46] developed the QDIA using a renormalized perturbation theory approach in which the isotropic problem was treated as the zero-order problem. The anisotropic and inhomogeneous terms were considered as small and multiplied by a perturbation parameter $\lambda$, prior to renormalization. Here we apply this approach to formulate the EDMAC.

The plan of this article is as follows. In Section 2, we present the equations for two-dimensional barotropic flows on a generalized $\beta$ – plane and with large-scale advection by a mean flow. The dynamical flow equations are then converted to the equivalent form in Fourier space in Section 3. This is necessary for the subsequent formulation, in Section 4, of non-Markovian closures for the interaction of anisotropic turbulence with Rossby waves. These generalized DIA, SCFT and LET closures have the same single-time cumulant equation and are equally suitable for reducing to the Markovian anisotropic closures (MACs) formulated in Section 5. There three variants of the MACs are formulated using the three versions of the FDT that are summarized in Equation (4). As well, in Section 5 the prognostic equations for the relaxation functions that close the Markovian statistical dynamical equations are derived. In Section 6, we focus on the Markovian variant for which the current-time FDT in Equation (1) is imposed and note the simplifications in the structure of the closure that then entail. The EDMAC model is formulated in Section 7 and in Section 8 sufficient conditions for the realizability of EDMAC in the presence of transient Rossby waves are established. Generalization of the realizable EDMAC to other physical systems with transient waves and to higher dimension and higher order nonlinearity is considered in Section 9. In Section 10, numerical integrations of the EDMAC and EDQNM models are reported on for isotropic and anisotropic 2D turbulence of various complexity. The implications of the results and our conclusions are presented in Section 11. The derivation, based on renormalized perturbation theory, of the generalized DIA closure from which our analysis starts is located in Appendix A. The Langevin equation for the EDMAC model that guarantees realizability is presented in Appendix B.

## 2. Two-dimensional Barotropic Flows on a $\beta$ –plane with Large-scale Advection

We formulate our statistical dynamical equations for the case of two-dimensional or barotropic flows, although the generalization to three-dimensional flows can also be accomplished as described in Section 9. Turbulent flows in planar geometry are considered on a generalized $\beta$ – plane and include a large-scale advection by a wind $U$. The total flow is described by the streamfunction $\Psi = \psi - Uy$ where $\psi$ represents the small scales. Throughout this paper, we present theoretical and numerical results for flows on the doubly periodic plane $0 \le x \le 2\pi,\ 0 \le y \le 2\pi$ with $\mathbf{x} = (x, y)$.

### *2.1 Large-scale flow equation*

The large-scale flow $U$ is kept the same in each realization of the smaller scales and effectively just modifies the beta effect. We include possible relaxation to a mean large-scale flow $U_0$ with strength $\alpha_U$ so that $U$ may evolve with time according to

$$\frac{\partial U}{\partial t} = \alpha_U (U_0 - U). \tag{5}$$

In the absence of the relaxation forcing and dissipative drag $U$ is separately conserved.

### *2.2 Barotropic vorticity equation for the small scales*

The evolution of the small scales is described by the barotropic vorticity equation on a $\beta$ – plane with the advective wind $U$ modifying the Rossby wave propagation:

$$\frac{\partial \zeta}{\partial t} = -J(\psi - Uy, \zeta + \beta y + k_0^2 U y) + \nu_0 \nabla^2 \zeta + f_0. \tag{6}$$

The Jacobian is defined by:

$$J(\psi, \zeta) = \frac{\partial \psi}{\partial x}\frac{\partial \zeta}{\partial y} - \frac{\partial \psi}{\partial y}\frac{\partial \zeta}{\partial x} \tag{7}$$

and the relationship between the vorticity $\zeta$ and the stream function $\psi$ is

$$\zeta = \nabla^2 \psi \equiv \left( \frac{\partial^2}{\partial x^2} + \frac{\partial^2}{\partial y^2} \right) \psi \tag{8}$$

where $\nabla^2$ is the Laplacian.

The results that we present can also easily be extended to turbulent flow on a sphere where $U$ is the analogue of the solid body rotation. Indeed, the structure of Equation (6) is the same as for flow on a sphere. The wavenumber $k_0$ is the analogue of that on the sphere with $k_0^2 U y$ corresponding to the vorticity of the solid body rotation. In Equation (6), $\nu_0$ is the viscosity, $\beta$ is the beta effect and $f_0$ specifies possible external forcing.

## 3. Dynamical Equations in Fourier Space

The statistical dynamics of turbulent flows is most conveniently analyzed in spectral space. For the doubly periodic domain the equations are transformed into Fourier space by spectral representations of the fields in the form:

$$\zeta(\mathbf{x},t) = \sum_{\mathbf{k}} \zeta_{\mathbf{k}}(t) \exp(i\,\mathbf{k.x}), \tag{9}$$

where the spectral coefficients

$$\zeta_{\mathbf{k}}(t) = \frac{1}{(2\pi)^2} \int_0^{2\pi} d^2\mathbf{x}\, \zeta(\mathbf{x},t) \exp(-i\,\mathbf{k.x}). \tag{10}$$

The wavenumbers in the $x$ and $y$ directions are $k_x$ and $k_y$ with the wavenumber vector $\mathbf{k} = (k_x, k_y)$ and the magnitude $k = (k_x^2 + k_y^2)^{1/2}$. The spectral coefficients satisfy the complex conjugate symmetry $\zeta_{-\mathbf{k}} = \zeta_{\mathbf{k}}^*$ which guarantees that the fields in physical space are real. The domain in Equation (9) can be general but we suppose that it contains the integer wavenumbers in a circular area (to be specified) that excludes the origin $\mathbf{0}$. The vorticity equation in spectral space then takes the form:

$$\begin{aligned} &\left(\frac{\partial}{\partial t} + \nu_0(k)k^2 + \mathrm{i}\omega_{\mathbf{k}}^U(t)\right)\zeta_{\mathbf{k}}(t) \\ &= \sum_{\mathbf{p}}\sum_{\mathbf{q}} \delta(\mathbf{k},\mathbf{p},\mathbf{q}) K(\mathbf{k},\mathbf{p},\mathbf{q}) \zeta_{-\mathbf{p}}(t)\zeta_{-\mathbf{q}}(t) + f_0(\mathbf{k},t). \end{aligned} \tag{11}$$

Here, we have generalized the form of the viscosity $\nu_0 \to \nu_0(k)$ which corresponds to more general dissipation operators than the Laplacian in Equation (6). In Equation (11) the delta function $\delta(\mathbf{k},\mathbf{p},\mathbf{q}) = 1$ if $\mathbf{k}+\mathbf{p}+\mathbf{q} = 0$ and 0 if $\mathbf{k}+\mathbf{p}+\mathbf{q} \neq 0$. The interaction coefficient $K(\mathbf{k},\mathbf{p},\mathbf{q})$ is defined by:

$$K(\mathbf{k},\mathbf{p},\mathbf{q}) = \tfrac{1}{2}[p_x q_y - p_y q_x](p^2 - q^2)/p^2 q^2. \tag{12}$$

The Doppler shifted Rossby wave frequency is given by the dispersion relationship:

$$\omega_{\mathbf{k}}^U(t) = \Omega_{\mathbf{k}}^U(t) + \omega_{\mathbf{k}}^{\beta} = \frac{U(t)k_x(k^2 - k_0^2)}{k^2} - \frac{\beta k_x}{k^2} \tag{13}$$

where the Rossby wave frequency is

$$\omega_{\mathbf{k}} \equiv \omega_{\mathbf{k}}^{\beta} = -\frac{\beta k_x}{k^2}, \tag{14}$$

and

$$\Omega_{\mathbf{k}}^U(t) = \frac{U(t)k_x(k^2 - k_0^2)}{k^2}. \tag{15}$$

We note that in the case of inviscid flows Rossby waves of the form $\exp \mathrm{i}(\mathbf{k.x} - \omega_{\mathbf{k}}^U t)$ are exact solutions to Equation (6).

## 4. Non-Markovian Closures for Turbulence and Rossby Waves

The aim with closure theory is to develop statistical dynamical equations that describe the evolution of the low order moments or cumulants for infinite ensembles of flow fields. To start the process, we first represent a given member $\zeta_{\mathbf{k}}(t)$ by its mean, the one-point cumulant, $<\zeta_{\mathbf{k}}(t)>\equiv\overline{\zeta}_{\mathbf{k}}(t)$, and the deviation from the mean, $\tilde{\zeta}_{\mathbf{k}}(t)$. We consider homogeneous turbulence in this study for which the small scales have zero mean:

$$<\zeta_{\mathbf{k}}(t)>\equiv\overline{\zeta}_{\mathbf{k}}(t)=0\ ;\ \zeta_{\mathbf{k}}(t)=\ \tilde{\zeta}_{\mathbf{k}}(t). \tag{16}$$

Since the mean is zero, the deviation $\tilde{\zeta}_{\mathbf{k}}$ also satisfies Equation (11) (with $\zeta_{\mathbf{k}}(t)\to\ \tilde{\zeta}_{\mathbf{k}}(t)$). The mean forcing also needs to be zero:

$$<\ f_0(\mathbf{k},t)>\equiv\ \overline{f}_0(\mathbf{k},t)=0\ ;\ f_0(\mathbf{k},t)=\ \tilde{f}_0(\mathbf{k},t)\ . \tag{17}$$

*4.1 Generalized DIA Closure for Anisotropic Turbulence and Rossby Waves*

To formulate the realizable EDMAC model equations we start with the DIA closure that is slightly generalized as derived in Appendix A. For homogeneous anisotropic turbulence the DIA closure describes the evolution of the two-time two-point cumulant

$$C_{\mathbf{k}}(t,t')=<\tilde{\zeta}_{\mathbf{k}}(t)\tilde{\zeta}_{-\mathbf{k}}(t')> \tag{18}$$

and response function

$$R_{\mathbf{k}}(t,t')=\left\langle\tilde{R}_{\mathbf{k}}(t,t')\right\rangle \tag{19}$$

which is the ensemble average of individual responses. The response function for an individual disturbance is

$$\tilde{R}_{\mathbf{k}}(t,t')=\frac{\delta\tilde{\zeta}_{\mathbf{k}}(t)}{\delta\,\tilde{f}_0(\mathbf{k},t')}. \tag{20}$$

The response function $\tilde{R}_{\mathbf{k}}(t,t')$ measures the change in the field $\delta\tilde{\zeta}_{\mathbf{k}}(t)$ due to an infinitesimal $\delta\,\tilde{f}_0(\mathbf{k},t')$ perturbation in the forcing at an earlier time. Here, $\delta$ denotes the functional derivative.

The two-time two-point cumulant equation for the generalized DIA, as formulated in Appendix A, is given by

$$\begin{aligned}&\left(\frac{\partial}{\partial t}+\nu_0(k)k^2+\mathrm{i}\omega^U_{\mathbf{k}}(t)\right)C_{\mathbf{k}}(t,t')\\&+\int_{t_0}^{t}ds\,(\eta_{\mathbf{k}}(t,s)+\pi^{\omega}_{\mathbf{k}}(t,s))C_{-\mathbf{k}}(t',s)\\&=\int_{t_0}^{t'}ds\,(S_{\mathbf{k}}(t,s)+P^{\omega}_{\mathbf{k}}(t,s)+F_0(\mathbf{k},t,s))R_{-\mathbf{k}}(t',s).\end{aligned} \tag{21}$$

This is the case for $t>t'$, while for $t'>t$ we have $C_{\mathbf{k}}(t,t')=C_{-\mathbf{k}}(t',t)$. Here, the last term in Equation (21) arises from

$$<\ \tilde{f}_0(\mathbf{k},t)\tilde{\zeta}_{-\mathbf{k}}(t')>=\int_{t_0}^{t'}ds\,F_0(\mathbf{k},t,s)R_{-\mathbf{k}}(t',s) \tag{22}$$

where $t_0$ is the initial time and the bare noise

$$F_0(\mathbf{k},t,s)=\ \ <\ \tilde{f}_0(\mathbf{k},t)\,\tilde{f}^*_0(\mathbf{k},s)>. \tag{23}$$

As well the nonlinear damping and nonlinear noise are derived in Appendix A and given by:

$$\eta_{\mathbf{k}}(t,s)=-4\sum_{\mathbf{p}}\sum_{\mathbf{q}}\delta(\mathbf{k},\mathbf{p},\mathbf{q})K(\mathbf{k},\mathbf{p},\mathbf{q})K(-\mathbf{p},-\mathbf{q},-\mathbf{k})R_{-\mathbf{p}}(t,s)C_{-\mathbf{q}}(t,s), \tag{24}$$

and

$$S_{\mathbf{k}}(t,s)=2\sum_{\mathbf{p}}\sum_{\mathbf{q}}\delta(\mathbf{k},\mathbf{p},\mathbf{q})K(\mathbf{k},\mathbf{p},\mathbf{q})K(-\mathbf{k},-\mathbf{p},-\mathbf{q})C_{-\mathbf{p}}(t,s)C_{-\mathbf{q}}(t,s). \tag{25}$$

The expressions for $\pi^{\omega}_{\mathbf{k}}(t,s)$ and $P^{\omega}_{\mathbf{k}}(t,s)$ are also derived in Appendix A and are given by:

$$\pi^{\omega}_{\mathbf{k}}(t,s) = R_{\mathbf{k}}(t,s)\omega^{U}_{\mathbf{k}}(t)\omega^{U}_{\mathbf{k}}(s) \tag{26}$$

and

$$P^{\omega}_{\mathbf{k}}(t,s) = C_{\mathbf{k}}(t,s)\omega^{U}_{\mathbf{k}}(t)\omega^{U}_{\mathbf{k}}(s). \tag{27}$$

In a similar way the response function equation can be derived as

$$\begin{aligned}&\left(\frac{\partial}{\partial t} + \nu_0(k)k^2 + \mathrm{i}\omega^{U}_{\mathbf{k}}(t)\right)R_{\mathbf{k}}(t,t') \\ &+\int_{t'}^{t} ds(\eta_{\mathbf{k}}(t,s) + \pi^{\omega}_{\mathbf{k}}(t,s))R_{\mathbf{k}}(s,t') = \delta(t-t')\end{aligned} \tag{28}$$

for $t \geq t'$ and the Dirac delta function means that $R_{\mathbf{k}}(t,t) = 1$.

The final equation needed for the DIA closure is that for the single-time two-point cumulant:

$$\begin{aligned}&\left(\frac{\partial}{\partial t} + 2\,\mathrm{Re}(\nu_0(k)k^2 + \mathrm{i}\omega^{U}_{\mathbf{k}}(t))\right)C_{\mathbf{k}}(t,t) \\ &+2\,\mathrm{Re}\int_{t_0}^{t} ds\,(\eta_{\mathbf{k}}(t,s) + \pi^{\omega}_{\mathbf{k}}(t,s))C_{-\mathbf{k}}(t,s) \\ &= 2\,\mathrm{Re}\int_{t_0}^{t} ds\,(S_{\mathbf{k}}(t,s) + P^{\omega}_{\mathbf{k}}(t,s) + F_0(\mathbf{k},t,s))R_{-\mathbf{k}}(t,s).\end{aligned} \tag{29}$$

The system of DIA equations is started from the initial conditions $C_{\mathbf{k}}(t_0,t_0)$ and $R_{\mathbf{k}}(t,t) = 1$. Equation (29) can also be simplified to the form:

$$\begin{aligned}&\left(\frac{\partial}{\partial t} + 2\nu_0(k)k^2\right)C_{\mathbf{k}}(t,t) + 2\,\mathrm{Re}\int_{t_0}^{t} ds\,\eta_{\mathbf{k}}(t,s)C_{-\mathbf{k}}(t,s) \\ &= 2\,\mathrm{Re}\int_{t_0}^{t} ds\,(S_{\mathbf{k}}(t,s) + F_0(\mathbf{k},t,s))R_{-\mathbf{k}}(t,s)\end{aligned} \tag{30}$$

since the $\pi^{\omega}_{\mathbf{k}}$ and $P^{\omega}_{\mathbf{k}}$ terms cancel because

$$\mathrm{Re}\{\pi^{\omega}_{\mathbf{k}}(t,s)C_{-\mathbf{k}}(t,s) - P^{\omega}_{\mathbf{k}}(t,s)R_{-\mathbf{k}}(t,s)\} = 0. \tag{31}$$

*4.2 The Abridged DIA Closure for Anisotropic Turbulence and Rossby Waves*

As a first step towards developing Markovian Anisotropic Closures (MACs) for the interaction of turbulence and Rossby waves we present in this subsection an abridged generalized DIA closure. We consider the situation where the large-scale field $U(t)$ is slowly varying in the time history integrals:

$$U(s) \to U(t) \;\; ; \;\; \omega^{U}_{\mathbf{k}}(s) \to \omega^{U}_{\mathbf{k}}(t). \tag{32}$$

Thus, the abridged DIA closure equations are again given by Equations (21) to (25) and (28) to (30) but with Equations (26) and (27) replaced by

$$\pi^{\omega}_{\mathbf{k}}(t,s) \to \pi^{\underline{\omega}}_{\mathbf{k}}(t,s) = R_{\mathbf{k}}(t,s)[\omega^{U}_{\mathbf{k}}(t)]^2, \tag{33}$$

and

$$P^{\omega}_{\mathbf{k}}(t,s) \to P^{\underline{\omega}}_{\mathbf{k}}(t,s) = C_{\mathbf{k}}(t,s)[\omega^{U}_{\mathbf{k}}(t)]^2, \tag{34}$$

where the Markov approximation in Equation (32) is denoted by the superscript $\underline{\omega}$.

*4.3 Generalized SCFT and LET Closures for Ansotropic Turbulence and Rossby Waves*

We can also arrive at generalized SCFT and LET closures from the generalized DIA closure in the usual way by invoking the prior-time FDT in Equation (2). For the generalized SCFT this FDT determines the two-time cumulant instead of Equation (21). For the generalized LET closure the FDT determines the response function instead of Equation (28). For the corresponding abridged SCFT and LET closures Equations (33) and (34) again replace Equations (26) and (27).

## 5. Statistical Dynamical Equations for Markovian Anisotropic Closures

*General Formulation of Markovian Anisotropic Closures*

As the next step towards formulating the realizable EDMAC model with analytical form for the relaxation function we outline the theory of associated Markovian Anisotropic Closures (MACs). The MACs are described by the single-time cumulant equation and auxiliary integral, or equivalent differential, equations for two relaxation functions. The MACS result from employing any of the three FDTs in Equation (4) and Markovianizing the generalized DIA in Section 4. As noted there the version with $X=0$ is the current-time FDT, $X=\frac{1}{2}$ the correlation FDT, and $X=1$ the prior-time FDT. The corresponding MACs are denoted by $MAC^X$. The MACs can be formulated using the general expressions for $\pi_{\mathbf{k}}^{\omega}(t,s)$ and $P_{\mathbf{k}}^{\omega}(t,s)$ in Equations (26) and (27). However, since our final aim is to simplify the MAC with $X=0$ to the realizable EDMAC we make the derivations from the abridged DIA with $\pi_{\mathbf{k}}^{\underline{\omega}}(t,s)$ and $P_{\mathbf{k}}^{\underline{\omega}}(t,s)$ given in Equations (33) and (34).

The single-time two-point cumulant equation for the abridged generalized DIA, (and as well for the corresponding SCFT and LET closure) that is also used for each of the MAC models can be written as:

$$\frac{\partial}{\partial t}C_{\mathbf{k}}(t,t)+2\,\mathrm{Re}[N_{\eta}(\mathbf{k},t)+N_{\underline{\omega}}(\mathbf{k},t)+N_0(\mathbf{k})] \\ =2\,\mathrm{Re}[F_S(\mathbf{k},t)+F_{\underline{\omega}}(\mathbf{k},t)+F_0(\mathbf{k},t)]. \tag{35}$$

We note that $C_{\mathbf{k}}(t,t)=C_{-\mathbf{k}}(t,t)$ is real. The $F_{\mathbf{k}}\ (t)$ and $N_{\mathbf{k}}\ (t)$ terms can be written in the following convenient forms for subsequent Markovianization. Firstly,

$$F_S(\mathbf{k},t)=2\sum_{\mathbf{p}}\sum_{\mathbf{q}}\delta(\mathbf{k},\mathbf{p},\mathbf{q})K(\mathbf{k},\mathbf{p},\mathbf{q})K(-\mathbf{k},-\mathbf{p},-\mathbf{q})\Delta(-\mathbf{k},-\mathbf{p},-\mathbf{q})(t), \tag{36}$$

where

$$\Delta(-\mathbf{k},-\mathbf{p},-\mathbf{q})(t)=\int_{t_0}^{t}dsR_{-\mathbf{k}}(t,s)C_{-\mathbf{p}}(t,s)C_{-\mathbf{q}}(t,s). \tag{37}$$

As well,

$$F_{\underline{\omega}}(\mathbf{k},t)=[\omega_{\mathbf{k}}^{U}(t)]^2\Lambda(-\mathbf{k},\mathbf{k})(t) \tag{38}$$

with

$$\Lambda(-\mathbf{k},\mathbf{k})(t)=\int_{t_0}^{t}dsR_{-\mathbf{k}}(t,s)C_{\mathbf{k}}(t,s). \tag{39}$$

The bare forcing spectrum

$$F_0(\mathbf{k},t)=\int_{t_0}^{t}dsF_0(\mathbf{k},t,s)R_{-\mathbf{k}}(t,s). \tag{40}$$

In a similar way,

$$N_{\eta}(\mathbf{k},t)=-4\sum_{\mathbf{p}}\sum_{\mathbf{q}}\delta(\mathbf{k},\mathbf{p},\mathbf{q})K(\mathbf{k},\mathbf{p},\mathbf{q})K(-\mathbf{p},-\mathbf{q},-\mathbf{k})\Delta(-\mathbf{p},-\mathbf{q},-\mathbf{k})(t), \tag{41}$$

with

$$N_{\underline{\omega}}(\mathbf{k},t)=[\omega_{\mathbf{k}}^{U}(t)]^{2}\Lambda(\mathbf{k},-\mathbf{k})(t), \tag{42}$$

and

$$N_{0}(\mathbf{k},t)=(\nu_{0}(k)k^{2}+\mathrm{i}\omega_{\mathbf{k}}^{U})C_{\mathbf{k}}(t,t)=D_{0}(\mathbf{k})C_{\mathbf{k}}(t,t). \tag{43}$$

Here, the dissipation and Rossby wave dispersion term is

$$D_{0}(\mathbf{k},t)=\nu_{0}(k)k^{2}+\mathrm{i}\omega_{\mathbf{k}}^{U}(t). \tag{44}$$

Next, the three versions of the FDT in Equation (4) are applied and this simplifies the nonlinear noise and nonlinear damping terms in Equations (36) and (41). Then the time history integrals can then be expressed by relaxation functions $\Theta^{X}$ and $\Psi^{X}$ for $X=0,\frac{1}{2},1$. The integral representations for $\Theta^{X}$ and $\Psi^{X}$ can in turn be replaced by differential equations that augment Equation (35) for the single-time cumulant. This system is therefore Markovian with the variant with $X=\frac{1}{2}$ also guaranteed to be realizable since it is a generalization of the RMC model of Bowman et al. [23].

From Equations (36), (37) and (4) we have

$$\begin{aligned}F_{S}(\mathbf{k},t)=2\sum_{\mathbf{p}}\sum_{\mathbf{q}}\delta(\mathbf{k},\mathbf{p},\mathbf{q})K(\mathbf{k},\mathbf{p},\mathbf{q})K(-\mathbf{k},-\mathbf{p},-\mathbf{q})\\ \times C_{-\mathbf{p}}^{1-X}(t,t)C_{-\mathbf{q}}^{1-X}(t,t)\Theta^{X}(-\mathbf{k},-\mathbf{p},-\mathbf{q})(t),\end{aligned} \tag{45}$$

with the triad relaxation function

$$\Theta^{X}(-\mathbf{k},-\mathbf{p},-\mathbf{q})(t)=\int_{t_0}^{t}dsR_{-\mathbf{k}}(t,s)R_{-\mathbf{p}}(t,s)R_{-\mathbf{q}}(t,s)C_{-\mathbf{p}}^{X}(s,s)C_{-\mathbf{q}}^{X}(s,s). \tag{46}$$

As well,

$$F_{\underline{\omega}}(\mathbf{k},t)=[\omega_{\mathbf{k}}^{U}(t)]^{2}C_{\mathbf{k}}^{1-X}\Psi^{X}(-\mathbf{k},\mathbf{k})(t) \tag{47}$$

where the relaxation function

$$\Psi^{X}(-\mathbf{k},\mathbf{k})(t)=\int_{t_0}^{t}dsR_{-\mathbf{k}}(t,s)R_{\mathbf{k}}(t,s)C_{\mathbf{k}}^{X}(s,s). \tag{48}$$

In a similar way, $N_{\eta}(\mathbf{k},t)$ and $N_{\underline{\omega}}(\mathbf{k},t)$ simplify to

$$N_{\eta}(\mathbf{k},t)=D_{\eta}(\mathbf{k},t)C_{\mathbf{k}}(t,t), \tag{49}$$

where

$$\begin{aligned}D_{\eta}(\mathbf{k},t)=-4\sum_{\mathbf{p}}\sum_{\mathbf{q}}\delta(\mathbf{k},\mathbf{p},\mathbf{q})K(\mathbf{k},\mathbf{p},\mathbf{q})K(-\mathbf{p},-\mathbf{q},-\mathbf{k})\\ \times C_{-\mathbf{q}}^{1-X}(t,t)C_{-\mathbf{k}}^{-X}(t,t)\Theta^{X}(-\mathbf{p},-\mathbf{q},-\mathbf{k})(t),\end{aligned} \tag{50}$$

and

$$N_{\underline{\omega}}(\mathbf{k},t)=D_{\underline{\omega}}(\mathbf{k},t)C_{\mathbf{k}}(t,t), \tag{51}$$

with

$$D_{\underline{\omega}}(\mathbf{k},t)=[\omega_{\mathbf{k}}^{U}(t)]^{2}C_{\mathbf{k}}^{-X}(t,t)\Psi^{X}(\mathbf{k},-\mathbf{k})(t). \tag{52}$$

The equation for the single-time cumulant can also be written in the form:

$$\left(\frac{\partial}{\partial t}+2\,\mathrm{Re}(D_{r}(\mathbf{k},t))\right)C_{\mathbf{k}}(t,t)=2\,\mathrm{Re}(F_{r}(\mathbf{k},t)) \tag{53}$$

and for consistency the response function equation becomes:

$$\frac{\partial}{\partial t}R_{\mathbf{k}}(t,t')+D_{r}(\mathbf{k},t)R_{\mathbf{k}}(t,t')=\delta(t-t'). \tag{54}$$

The renormalized dissipation and noise functions in Equations (53) and (54) are defined by

$$D_r(\mathbf{k},t) = D_0(\mathbf{k}) + D_\eta(\mathbf{k},t) + D_{\underline{\omega}}(\mathbf{k},t); \\ F_r(\mathbf{k},t) = F_0(\mathbf{k},t) + F_S(\mathbf{k},t) + F_{\underline{\omega}}(\mathbf{k},t) \quad (55)$$

with the contributing terms given above.

The modified form of the response function can also be written as:

$$\left(\frac{\partial}{\partial t} + \nu_0(k)k^2 + \mathrm{i}\omega_{\mathbf{k}}^U(t)\right) R_{\mathbf{k}}(t,t') - \delta(t-t') \\ = -\left[\int_{t_0}^{t} ds(\eta_{\mathbf{k}}^X(t,s) + \pi_{\mathbf{k}}^{\omega}(t,s)) R_{-\mathbf{k}}(t,s)[C_{\mathbf{k}}(s,s)]^X [C_{\mathbf{k}}(t,t)]^{-X}\right] R_{\mathbf{k}}(t,t') \quad (56)$$

for $t \geq t'$ with $R_{\mathbf{k}}(t,t) = 1$ and $C_{\mathbf{k}}(t,t) = C_{-\mathbf{k}}(t,t)$ is real. Here we have used the general FDT in Equation (4) and $\eta_{\mathbf{k}}^X$ is then also obtained from $\eta_{\mathbf{k}}$ in Equation (24) and has the expression:

$$\eta_{\mathbf{k}}^X(t,s) = -4\sum_{\mathbf{p}}\sum_{\mathbf{q}} \delta(\mathbf{k},\mathbf{p},\mathbf{q}) K(\mathbf{k},\mathbf{p},\mathbf{q}) K(-\mathbf{p},-\mathbf{q},-\mathbf{k}) \\ \times R_{-\mathbf{p}}(t,s) R_{-\mathbf{q}}(t,s) C_{\mathbf{q}}^{1-X}(t,t) C_{\mathbf{q}}^X(s,s). \quad (57)$$

The simplified form of the response function in Equation (54) together with the FDT in Equation (4) means that the relaxation functions $\Theta^X$ in Equation (46) and $\Psi^X$ in Equation (48) can be replaced by differential equations. The differential equation for $\Theta^X$ is:

$$\frac{\partial}{\partial t}\Theta^X(\mathbf{k},\mathbf{p},\mathbf{q})(t) + (D_r(\mathbf{k},t) + D_r(\mathbf{p},t) + D_r(\mathbf{q},t))\Theta^X(\mathbf{k},\mathbf{p},\mathbf{q})(t) \\ = C_{\mathbf{p}}^X(t,t) C_{\mathbf{q}}^X(t,t) \quad (58)$$

where $\Theta^X(\mathbf{k},\mathbf{p},\mathbf{q})(0) = 0$ and $D_{\mathbf{k}}^r$ is given in Equation (55). The corresponding differential equation for $\Psi^X$ is:

$$\frac{\partial}{\partial t}\Psi^X(\mathbf{k},-\mathbf{k})(t) + (D_r(\mathbf{k},t) + D_r(-\mathbf{k},t))\Psi^X(\mathbf{k},-\mathbf{k})(t) = C_{\mathbf{k}}^X(t,t) \quad (59)$$

with $\Psi^X(\mathbf{k},-\mathbf{k})(0) = 0$.

Now, the closure for any of the three MACs with $X = 0,\ \frac{1}{2},\ 1$ is Equation (53) for the single-time cumulant $C_{\mathbf{k}}(t,t)$, augmented by the differential equations for $\Theta^X(\mathbf{k},\mathbf{p},\mathbf{q})(t)$ in Equation (58) and $\Psi^X(\mathbf{k},-\mathbf{k})(t)$ in Equation (59). The replacement of the integral forms for $\Theta^X$ and $\Psi^X$ by the differential equations has made the system Markovian. The auxiliary differential equations for $\Theta^X$ and $\Psi^X$ of course generate the same information as the time history integral forms but for long time simulations are more efficient. Nevertheless, analytical forms for $\Theta^X$ and $\Psi^X$ result in even more efficient closures such as the EDQNM and EDMAC of Section 7. Each of the three $MAC^X$ formulated in this Section is realizable for pure turbulence without transient wave phenomena. The MAC model with $X = \frac{1}{2}$ is also realizable when transient Rossby waves are present [23]. We note that in fact

$$D_{\underline{\omega}}(\mathbf{k},t) C_{\mathbf{k}}(t,t) = F_{\underline{\omega}}(\mathbf{k},t) \quad (60)$$

so these terms can be dropped from Equation (53).

## 6. Markovian Anisotropic Closure with Current-Time FDT

In this Section, we consider the further simplifications of the $MAC^X$ models when $X=0$ so that the current-time FDT is used. This is a step towards the formulation of the EDMAC model, in the following Section, with analytical forms for the relaxation functions. When $X=0$, Equation (53) for the single-time cumulant simplifies to:

$$\begin{aligned}&\left(\frac{\partial}{\partial t}+2\nu_0(k)k^2\right)C_{\mathbf{k}}(t,t)=N_{\mathbf{k}}(t)+2\mathrm{Re}F_0(\mathbf{k},t)\\&=8\sum_{\mathbf{p}}\sum_{\mathbf{q}}\delta(\mathbf{k},\mathbf{p},\mathbf{q})K(\mathbf{k},\mathbf{p},\mathbf{q})K(\mathbf{p},\mathbf{q},\mathbf{k})\,\mathrm{Re}\,\Theta^{X=0}(\mathbf{k},\mathbf{p},\mathbf{q})(t)\\&\quad\times C_{\mathbf{q}}(t,t)\left[C_{\mathbf{k}}(t,t)-C_{\mathbf{p}}(t,t)\right]+2\mathrm{Re}F_0(\mathbf{k},t),\end{aligned}\tag{61}$$

where, for white noise $2\,\mathrm{Re}F_0(\mathbf{k},t)=F_0(\mathbf{k},t,t)$. Here, we have used Equation (60) and the fact that the interaction coefficients have the properties that $K(\mathbf{k},\mathbf{p},\mathbf{q})=K(\mathbf{k},\mathbf{q},\mathbf{p})$, and $K(\mathbf{k},\mathbf{p},\mathbf{q})+K(\mathbf{p},\mathbf{q},\mathbf{k})+K(\mathbf{q},\mathbf{k},\mathbf{p})=0$. Also note that single-time cumulants are real. As well, to establish Equation (61) we have used the fact that the triad relaxation function $\Theta^{X=0}$ is symmetric in the three wavenumber indices:

$$\Theta^{X=0}(\mathbf{k},\mathbf{p},\mathbf{q})(t)=\int_{t_0}^{t}ds R_{\mathbf{k}}(t,s)R_{\mathbf{p}}(t,s)R_{\mathbf{q}}(t,s)\tag{62}$$

as seen from Equation (46) when $X=0$.

Similarly, the expression for $\Psi^{X=0}$ simplifies to

$$\Psi^{X=0}(\mathbf{k},-\mathbf{k})(t)=\Psi^{X=0}(-\mathbf{k},\mathbf{k})(t)=\int_{t_0}^{t}ds R_{-\mathbf{k}}(t,s)R_{\mathbf{k}}(t,s).\tag{63}$$

The corresponding differential equation forms of $\Theta^{X=0}$ and $\Psi^{X=0}$ are given in Equations (58) and (59) with the simplification that $C^{X=0}\to 1$ on the right-hand sides. Also, when $X=0$, Equations (50) reduces to

$$D_\eta(\mathbf{k},t)=-4\sum_{\mathbf{p}}\sum_{\mathbf{q}}\delta(\mathbf{k},\mathbf{p},\mathbf{q})K(\mathbf{k},\mathbf{p},\mathbf{q})K(\mathbf{p},\mathbf{q},\mathbf{k})C_{\mathbf{q}}(t,t)\Theta^{X=0}(\mathbf{k},\mathbf{p},\mathbf{q})(t)\tag{64}$$

where we have used the properties of the interaction coefficients noted above and the symmetry properties of the triad relaxation function. As well, Equation (52) becomes

$$D_{\varpi}(\mathbf{k},t)=[\omega_{\mathbf{k}}^U(t)]^2\Psi^{X=0}(\mathbf{k},-\mathbf{k})(t)=[\omega_{\mathbf{k}}^U(t)]^2\int_{t_0}^{t}ds R_{\mathbf{k}}(t,s)R_{-\mathbf{k}}(t,s).\tag{65}$$

We also note that if the current-time FDT in Equation (1) is used (with $X=0$ in Equation (4)) then the expression for the response function simplifies to

$$\begin{aligned}&\left(\frac{\partial}{\partial t}+\nu_0(k)k^2+\mathrm{i}\omega_{\mathbf{k}}^U(t)\right)R_{\mathbf{k}}(t,t')\\&+\left[\int_{t_0}^{t}ds(\eta_{\mathbf{k}}^{X=0}(t,s)+[\omega_{\mathbf{k}}^U(t)]^2R_{\mathbf{k}}(t,s))R_{-\mathbf{k}}(t,s)\right]R_{\mathbf{k}}(t,t')\\&=\delta(t-t')\end{aligned}\tag{66}$$

for $t\ge t'$ with $R_{\mathbf{k}}(t,t)=1$ and $\eta_{\mathbf{k}}^X$ is defined in Equation (57). Here we note that the expression for $D_\eta(\mathbf{k},t)$ in Equation (64) where $X=0$ can also be written as

$$D_\eta(\mathbf{k},t)=\int_{t_0}^{t}ds\eta_{\mathbf{k}}^{X=0}(t,s)R_{-\mathbf{k}}(t,s)\tag{67}$$

and $D_{\varpi}(\mathbf{k},t)$ is given in Equation (52) with $C^{X=0}\to 1$. Thus, Equation (66) is of course just Equation (54) written out in detail when $X=0$.

For general $X = 0,\ \frac{1}{2},\ 1$, corresponding to the three FDTs, the solution to the response function differential Equation (54) (and Equation (66)) is

$$R_{\mathbf{k}}(t,t') = \exp\left(-\int_{t'}^{t} ds D_r(\mathbf{k},s)\right) \tag{68}$$

where $D_r$ is defined in Equation (55). Of course, when $X = 0$, $D_\eta$ and $D_{\underline{\omega}}$ are given in Equations (64) and (65). We also note that, for $X = 0$ the triad relaxation function in Equation (46) reduces to

$$\Theta^{X=0}(\mathbf{k},\mathbf{p},\mathbf{q})(t) = \int_{t_0}^{t} dt' \exp\left(-\int_{t'}^{t} ds\left[D_r(\mathbf{k},s)+D_r(\mathbf{p},s)+D_r(\mathbf{q},s)\right]\right). \tag{69}$$

**7. Realizable Eddy-Damped Markovian Anisotropic Closure**

In this Section, we consider further simplifications that lead to the realizable EDMAC model. The approach involves the Markovian approximation for the damping terms, as in the EDQNM, and leads to analytical expressions for the relaxation functions.

*7.1 Markovian Approximation and Analytical Eddy Damping*

Thus, with the Markov approximation for $D_r$, namely $D_r(\mathbf{k},s) \to D_r(\mathbf{k},t)$, the response function takes the simple form:

$$R_{\mathbf{k}}(t,t') = \exp\left(-D_r(\mathbf{k},t)(t-t')\right), \tag{70}$$

since integral in Equation (68) can then be evaluated. As well, the integrals in the expression for the triad relaxation function $\Theta^{X=0}$ in Equation (69) can also be evaluated to give

$$\begin{aligned}\Theta^{X=0}(\mathbf{k},\mathbf{p},\mathbf{q})(t) &= \int_{t_0}^{t} dt' \exp\left(-\left[D_r(\mathbf{k},t)+D_r(\mathbf{p},t)+D_r(\mathbf{q},t)\right](t-t')\right) \\ &= \frac{1-\exp\left(-\left[D_r(\mathbf{k},t)+D_r(\mathbf{p},t)+D_r(\mathbf{q},t)\right](t-t_0)\right)}{\left[D_r(\mathbf{k},t)+D_r(\mathbf{p},t)+D_r(\mathbf{q},t)\right]}.\end{aligned} \tag{71}$$

For the EDMAC model, like the EDQNM closure, we use a parameterized form for the eddy damping $D_\eta$, which appears in Equation (71) through $D_r = D_0 + D_\eta + D_{\underline{\omega}}$, and that is consistent with the $k^{-3}$ forward enstrophy cascading inertial range:

$$D_\eta(\mathbf{k},t) = \int_{t_0}^{t} ds \eta_{\mathbf{k}}^{X=0}(t,s) R_{-\mathbf{k}}(t,s) \to \mu_{\mathbf{k}}^{eddy}(t) = \gamma\left[k^2 C_{\mathbf{k}}(t,t)\right]^{\frac{1}{2}} \tag{72}$$

where

$$\mu_{\mathbf{k}}(t) = \nu_0(k)k^2 + \mu_{\mathbf{k}}^{eddy}(t) = \hat{\nu}_0(k)k^2 + \gamma\left[k^2 C_{\mathbf{k}}(t,t)\right]^{\frac{1}{2}}. \tag{73}$$

Here $\gamma$ is a positive dimensionless parameter. This local form for $\mu_{\mathbf{k}}^{eddy}(t)$ is consistent with that of Leith [2] for 2D turbulence, and Orszag [1] used a corresponding local form for 3D turbulence applicable to the $k^{-\frac{5}{3}}$ forward energy cascading inertial range. Note, however, that our approach applies equally if an integral form over wavenumbers (Pouquet et al. [50]; Herring et al. [51]) is used instead for the eddy damping.

*7.2 Frequency-Dependent Damping from Renormalized Perturbation Theory*

Here, we consider the problem where the damping $\mu_{\mathbf{k}}(t)$ is taken as the zero-order term and the Doppler shifted frequency $\omega_{\mathbf{k}}^{U}(t) \to \lambda\omega_{\mathbf{k}}^{U}(t)$ is supposed to be a small perturbation of order $\lambda$. Then, we carry out the perturbation expansion as in Appendix A with the result that:

$$\left(\frac{\partial}{\partial t}+\mu_{\mathbf{k}}(t)\right)R_{\mathbf{k}}(t,t')-\delta(t-t')$$
$$=-\mathrm{i}\lambda\omega_{\mathbf{k}}^{U}(t)R_{\mathbf{k}}^{(0)}(t,t')-\lambda^{2}\omega_{\mathbf{k}}^{U}(t)\left[\int_{t_0}^{t}ds\,\omega_{\mathbf{k}}^{U}(s)R_{\mathbf{k}}^{(0)}(t,s)R_{-\mathbf{k}}^{(0)}(t,s)\right]R_{\mathbf{k}}^{(0)}(t,t') \tag{74}$$
$$\approx-\mathrm{i}\lambda\omega_{\mathbf{k}}^{U}(t)R_{\mathbf{k}}^{(0)}(t,t')-\lambda^{2}[\omega_{\mathbf{k}}^{U}(t)]^{2}\left[\int_{t_0}^{t}ds\,R_{\mathbf{k}}^{(0)}(t,s)R_{-\mathbf{k}}^{(0)}(t,s)\right]R_{\mathbf{k}}^{(0)}(t,t').$$

Here, $R_{\mathbf{k}}^{(0)}(t,t')=\exp-\mu_{\mathbf{k}}(t)(t-t')$ is the zero-order response function as in Appendix A. We have assumed that $\mu$ and $\omega$ are slowly varying and then we can evaluate the integral to obtain:

$$\left(\frac{\partial}{\partial t}+\mu_{\mathbf{k}}(t)\right)R_{\mathbf{k}}(t,t')-\delta(t-t')$$
$$\approx-\mathrm{i}\lambda\omega_{\mathbf{k}}^{U}(t)R_{\mathbf{k}}^{(0)}(t,t')-\lambda^{2}\frac{[\omega_{\mathbf{k}}^{U}(t)]^{2}}{2\mu_{\mathbf{k}}(t)}\left[1-\exp-(2\mu_{\mathbf{k}}(t)t)\right]R_{\mathbf{k}}^{(0)}(t,t'). \tag{75}$$

In formulating the EDQNM model $\mu_{\mathbf{k}}^{eddy}(t)$ is specified to have the form in Equation (72) (or the integral form of Pouquet et al. [50]) from the initial conditions rather than developing from zero. In the same way, we take the quasi-steady state expression for the contribution from $[\omega_{\mathbf{k}}^{U}(t)]^{2}$. The result is:

$$\left(\frac{\partial}{\partial t}+\mu_{\mathbf{k}}(t)\right)R_{\mathbf{k}}(t,t')$$
$$\approx-\mathrm{i}\lambda\omega_{\mathbf{k}}^{U}(t)R_{\mathbf{k}}^{(0)}(t,t')-\lambda^{2}\frac{[\omega_{\mathbf{k}}^{U}(t)]^{2}}{2\mu_{\mathbf{k}}(t)}R_{\mathbf{k}}^{(0)}(t,t')+\delta(t-t'). \tag{76}$$

Renormalizing, we obtain our required expression for establishing the realizable EDMAC model:

$$\left(\frac{\partial}{\partial t}+\mu_{\mathbf{k}}(t)\right)R_{\mathbf{k}}(t,t')$$
$$\approx-\mathrm{i}\omega_{\mathbf{k}}^{U}(t)R_{\mathbf{k}}(t,t')-\frac{[\omega_{\mathbf{k}}^{U}(t)]^{2}}{2\mu_{\mathbf{k}}(t)}R_{\mathbf{k}}(t,t')+\delta(t-t'). \tag{77}$$

Thus, comparing the frequency-dependent term $D_{\underline{\omega}}$, for the MAC model, in Equation (65) with Equation (77) we see that

$$D_{\underline{\omega}}(\mathbf{k},t)\to c\frac{[\omega_{\mathbf{k}}^{U}(t)]^{2}}{\mu_{\mathbf{k}}(t)} \tag{78}$$

where the above analysis indicates that $c=\frac{1}{2}$. We leave the constant $c$ in Equation (78) and in the following analysis since it could be used as an empirical parameter in situations when the wave frequency is not small compared with the eddy damping. From Equation (52), with $C^{X=0}\to 1$, and Equation (78) we see that

$$\Psi^{EDMAC}(\mathbf{k},-\mathbf{k})(t)=\frac{c}{\mu_{\mathbf{k}}(t)} \tag{79}$$

where $\mu_{\mathbf{k}}(t)$ is given in Equation (73).

Combining the above results with those of Subsection 7.1, we find

$$D_{r}(\mathbf{k},t)\to\rho_{\mathbf{k}}(t)+\mathrm{i}\omega_{\mathbf{k}}^{U}(t) \tag{80}$$

where

$$\rho_{\mathbf{k}}(t) = \mu_{\mathbf{k}}(t) + c\frac{[\omega_{\mathbf{k}}^U(t)]^2}{\mu_{\mathbf{k}}(t)} > 0 \tag{81}$$

is the frequency renormalized damping. Thus, the EDMAC response function becomes:

$$R_{\mathbf{k}}(t,t') \approx R_{\mathbf{k}}^{EDMAC}(t,t') = \exp\left(-[\rho_{\mathbf{k}}(t) + \mathrm{i}\omega_{\mathbf{k}}^U(t)](t-t')\right), \tag{82}$$

and the EDMAC triad relaxation function:

$$\begin{aligned}&\Theta^{EDMAC}(\mathbf{k},\mathbf{p},\mathbf{q})(t)\\ &= \frac{1-\exp\left(-\left[\rho_{\mathbf{k}}(t)+\rho_{\mathbf{p}}(t)+\rho_{\mathbf{q}}(t)+\mathrm{i}(\omega_{\mathbf{k}}^U(t)+\omega_{\mathbf{p}}^U(t)+\omega_{\mathbf{q}}^U(t))\right](t-t_0)\right)}{\rho_{\mathbf{k}}(t)+\rho_{\mathbf{p}}(t)+\rho_{\mathbf{q}}(t)+\mathrm{i}(\omega_{\mathbf{k}}^U(t)+\omega_{\mathbf{p}}^U(t)+\omega_{\mathbf{q}}^U(t))}.\end{aligned} \tag{83}$$

The EDMAC model is then defined by Equation (61), but with $\Theta^{EDMAC}$ in Equation (83) replacing $\Theta^{X=0}$. In Section 8, to follow, and Appendix B, we determine sufficient conditions for the EDMAC to be realizable. The EDMAC has the same structure as the EDQNM but with just $\rho_{\mathbf{k}}$ replacing $\mu_{\mathbf{k}}$ in the triad relaxation function. It is essentially as computationally efficient as the EDQNM, as discussed further in Section 10, but with the advantage of guaranteed realizability.

## 8. Conditions for Realizability of EDMAC with Variable Rossby Wave Frequency

In this Section, we determine sufficient conditions on the damping $\rho_{\mathbf{k}}(t) > 0$, in Equation (81), so that $\mathrm{Re}\,\Theta^{EDMAC} \geq 0$; that is, so that the real part of the triad relaxation in Equation (83) is positive semi-definite. We consider the general case where the Doppler shifted frequencies $\omega_{\mathbf{k}}^U(t)$ are time dependent and are not necessarily monotonic.

The results of Section 7 suggest that $c = \frac{1}{2}$ is an appropriate value for small $\omega_{\mathbf{k}}^U(t)$. More generally, for larger $\omega_{\mathbf{k}}^U(t)$ the constant $c$ may be regarded as an empirical factor that is specified. Our aim is therefore to determine sufficient conditions on $c$ for realizability of the EDMAC model.

The triad relaxation function, $\Theta^{EDMAC}(t) = \Theta^{EDMAC}(\mathbf{k}_1,\mathbf{k}_2,\mathbf{k}_3)(t)$, in Equation (83) for the EDMAC model can be written in simplified form as:

$$\Theta^{EDMAC}(t) = \frac{1-\exp-[\rho+\mathrm{i}\omega\ ](t-t_0)}{\rho+\mathrm{i}\omega} \tag{84}$$

where we take $t_0 = 0$ (without loss of generality). In Equation (84),

$$\begin{aligned}\mu(t) &= \sum_{j=1}^{N}\mu_{\mathbf{k}_j}(t) > 0,\\ \omega(t) &= \sum_{j=1}^{N}\omega_{\mathbf{k}_j}^U(t),\\ \rho(t) &= \sum_{j=1}^{N}\rho_{\mathbf{k}_j}(t) = \sum_{j=1}^{N}\left(\mu_{\mathbf{k}_j}(t) + c\frac{[\omega_{\mathbf{k}_j}^U(t)]^2}{\mu_{\mathbf{k}_j}(t)}\right) > 0,\end{aligned} \tag{85}$$

with $N = 3$ for triad interactions. The real part of the relaxation function that appears in the EDMAC model for the two-point cumulant (Equations (53) and (61) with superscript $X = 0$ replaced by $EDMAC$) is then given by:

$$\begin{aligned}&\mathrm{Re}\,\Theta^{EDMAC}(t)\\ &= \frac{1}{\rho^2+\omega^2}\left[\rho\{1-[\exp-\rho t]\cos\omega t\} + \omega[\exp-\rho t]\sin\omega t\right].\end{aligned} \tag{86}$$

Then, sufficient conditions for $\mathrm{Re}\,\Theta^{EDMAC}(t) \geq 0$ are determined as follows. In Equation (86) we can replace $\omega$ by $|\omega|$ for $t \geq 0$, since both $\cos\omega t$ and $\omega\sin\omega t$ are even functions of $\omega$. On the basis of the analysis in Section 7.2, we suppose that

$$\rho \geq |\omega|. \tag{87}$$

Firstly, we see that $\mathrm{Re}\,\Theta^{EDMAC}(t) \geq 0$, for $0 < |\omega(t)|\, t \leq \pi$ since then

$$1-[\exp-\rho t]\cos|\omega|t > 0 \;\; ; \;\; |\omega| \; [\exp-\rho t]\sin|\omega|t \geq 0. \tag{88}$$

Secondly, from Equation (87), we see that for $|\omega(t)|\,t > \pi$,

$$\exp-\rho t \leq \exp-(|\omega|t) < \exp-(\pi) < 0.05. \tag{89}$$

Further, in general

$$-1 \leq \cos|\omega|t \leq 1; \quad -1 \leq \sin|\omega|t \leq 1 \tag{90}$$

and thus for $|\omega(t)|\,t > \pi$,

$$\left[\rho^2 + \omega^2\right]\mathrm{Re}\,\Theta^{EDMAC}(t) > 0.9\rho > 0. \tag{91}$$

This then means that $\mathrm{Re}\,\Theta^{EDMAC}(t) \geq 0$, for $0 \leq |\omega(t)|\,t < \infty$. Sufficient conditions for the validity of Equation (87) is that each wavevector component satisfies

$$\rho_{\mathbf{k}_j}(t) = \left(\mu_{\mathbf{k}_j}(t) + c\frac{[\omega^U_{\mathbf{k}_j}(t)]^2}{\mu_{\mathbf{k}_j}(t)}\right) \geq |\omega^U_{\mathbf{k}_j}(t)|. \tag{92}$$

In turn, the inequalities in Equation (92) are valid provided $c \geq \frac{1}{4}$ as seen by solving quadratic equations.

## 9. Generalizations of the EDMAC Model

The analysis of Section 8, determining sufficient conditions for $\mathrm{Re}\,\Theta^{EDMAC} \geq 0$, can in fact be generalized to all dimensions $d \geq 2$ and with $N \geq 3$ components interacting instead of just the three. The condition for this is again that $c \geq \frac{1}{4}$ since none of the argument in Section 8 depends on the dimension or the number of interacting components. It just requires that $\mu_{\mathbf{k}}(t) > 0$ and $\omega^U_{\mathbf{k}}(t)$, with a general dispersion relationship, is finite.

This again has important implications for general EDQNM-type closures for which realizability is dependent on the relaxation function having positive semi-definite real part. Suppose that the relaxation function $\Theta^{EDQNM}(t)$ is given by the right hand side of Equation (84) but with $\mu$ replacing $\rho$. Then including the effects of frequency renormalized damping, $\mu \to \rho$, as specified in Equation (85) results in the associated realizable EDMAC model.

It is straightforward to extend the EDMAC model to 2D turbulent flows and Rossby waves on a differentially rotating sphere [9,27]. It is also a simple matter to extend the EDMAC to 3D quasigeostrophic turbulent flows and waves in the model in Appendix B of Frederiksen [48] generalized to rotating flows on a $\beta$–plane and on a sphere. The EDQNM closure of Carnevale and Frederiksen [25] for two-dimensional internal gravity waves used the quasi-steady state form for the triad relaxation function to ensure realizability. This restriction can be removed, and transient internal gravity waves considered, by going to the EDMAC model form for the relaxation functions. Realizability can be assured by using the frequency renormalized form of the damping in Equation (81) where $\omega^U_{\mathbf{k}}(t)$ is replaced by the internal wave frequency $\omega^s_{\mathbf{k}}(t) = sk_x/k$ with $s = \pm 1$ (Equation (2.7c) of Ref. [25]).

One might also expect to be able to develop suitable generalizations of the EDMAC to 3D Navier Stokes turbulence [1,10,22], including with rotation and waves. In particular, the study by Cambon and Jacqui [52] considered 3D anisotropic turbulence subject to rotation with different Rossby numbers and waves in EDQNM type models. They also used the quasi-steady state form for the triad relaxation function that ensures realizability. This might be extended to the transient regime form by using a frequency renormalized damping as for the EDMAC model. Rose and Sulem [53] and Clark et al, [54] formulate the EDQNM closure for isotropic turbulence in general $d \geq 2$ dimensions and it would be of interest to see if this could be generalized to anisotropic turbulence, with preferred zonal motion [27,36,37], subject to differential rotation and waves with the realizable EDMAC.

There are of course also important physical systems for which the nonlinearity is of higher order than quadratic. Two examples, with cubic nonlinearity in the field equations, are the nonlinear Schrodinger equation (Nazarenko [55] reviews the literature) and the Klein Gordon equation with $\lambda\phi^4$ Lagrangian (Frederiksen [49] reviews the literature). The nonlinear Schrodinger equation is of major importance in the study of optical turbulence, plasma physics and continuum mechanics [55,56] and the Klein Gordon equation in the classical and quantum statistical dynamics of Bose-Einstein condensation and scalar field theory [49,57,58]. In spectral space, $N = 4$ in Equation (85), corresponding to quartic interactions, would then be needed in closures including EDMAC, EDQNM and resonant interaction or wave turbulence models [25,49,54–58] of such phenomena.

## 10. Comparison of Closure Integrations for Turbulence and Rossby Wave Dynamics

In this Section, we compare closure calculations with the EDQNM and EDMAC models for isotropic turbulence and anisotropic turbulence with Rossby waves and possibly a large-scale mean flow. The integrations start from the isotropic spectrum B that was studied by Frederiksen and Davies [15,16] for closures with discrete spectra and which is very similar to the continuous spectrum II of Herring et al. [59]. These spectra were used by Herring et al. [59] and and Frederiksen and Davies [15] in studies of the comparison of closures with DNS for 2D isotropic turbulence. In particular, the continuous wavenumber formulation of the DIA closure was compared with discrete wavenumber DNS by Herring et al. [59] while Frederiksen and Davies [15] used the discrete formulation for both. In both studies the DIA closure, and associated SCFT and LET closures in Ref. [15], were found to underestimate the small-scale amplitudes (Figures 22 and 23 of Ref. [59]; Figures 3 and 4 of Ref. [15]). However, the discrete closures [15] were found to be in better agreement with the statistics of DNS than the continuous closures [59]. O'Kane and Frederiksen [17] also used spectrum B [15,16] for the initial transient spectrum in studies of inhomogeneous turbulence interacting with mean flows and topography with the QDIA closure.

A regularized version of the DIA, the RDIA, in which the wavenumber ranges of interaction are restricted in the response function and two-time cumulant [16], was found to give quite close agreement with the statistics of DNS at the expense of an empirical parameter $\alpha$. This is shown in Figures 1 and 2 of Ref. [16], although there does seem to be a tendency of the smallest scales of the DNS to kick up somewhat in all these studies [15,16,59].

The EDQNM and EDMAC models also depend on the empirical parameter $\gamma$ in Equation (72). For the EDMAC model we have estimated $c = \frac{1}{2}$ to be the strength of the frequency-dependent damping in Section 7.1 and we use that value. More generally, if the waves do not satisfy the formal requirement of being of small amplitude, then $c$ could also be regarded as an empirical parameter. Here, our aim is first to compare the EDQNM and EDMAC models for a commonly used value of $\gamma = 0.6$ [2,9] for 2D turbulence and then to discuss the generality of our findings for a range of $\gamma$.

### *10.1 Diagnostics*

The evolved closure integrations for the EDQNM and EDMAC models are compared in terms of the similarity of their evolved kinetic energy and palinstrophy spectra and Reynolds number and skewness. These diagnostics are defined as follows. Firstly, we specify the energy

$$E = \tfrac{1}{2}\sum_{\mathbf{k}} C_{\mathbf{k}}(t,t)k^{-2}, \tag{93}$$

enstrophy

$$F = \tfrac{1}{2}\sum_{\mathbf{k}} C_{\mathbf{k}}(t,t), \tag{94}$$

palinstrophy

$$P = \tfrac{1}{2}\sum_{\mathbf{k}} k^2 C_{\mathbf{k}}(t,t), \tag{95}$$

enstrophy dissipation

$$\eta = \sum_{\mathbf{k}} \nu_0 k^2 C_{\mathbf{k}}(t,t) = 2\nu_0 P\ , \tag{96}$$

and palintrophy production

$$K = \sum_{\mathbf{k}} k^2 N_{\mathbf{k}}(t). \tag{97}$$

Here, $N_{\mathbf{k}}(t)$ is defined in Equation (61). The large-scale Reynolds number and skewness are then specified by

$$R_L(t) = E \Big/ (\nu_0 \eta^{\frac{1}{3}}) \tag{98}$$

and

$$S(t) = 2K \Big/ (PF^{\frac{1}{2}}). \tag{99}$$

The transient kinetic energy spectra, and palinstrophy spectra, averaged over circular bands, are defined by:

$$E(k_i,t) = \tfrac{1}{2}\sum_{\mathbf{k}\in S} C_{\mathbf{k}}(t,t)k^{-2}, \tag{100}$$

and

$$P(k_i,t) = \tfrac{1}{2}\sum_{\mathbf{k}\in S} k^2 C_{\mathbf{k}}(t,t), \tag{101}$$

with the set $S$ is defined by

$$S = [\mathbf{k} \mid k_i = \mathrm{Int.}(|\mathbf{k}| + \tfrac{1}{2})]. \tag{102}$$

The band-average is over all $\mathbf{k}$ within a band of unit width at a radius $k_i$.

*10.2 Initial Spectra and Parameter Specifications*

In this subsection, we specify the parameters and initial spectra used in our closure calculations. They are based on the related studies in Refs. [9,15,16,47,59]. The length scale used is of one half the earth's radius, $a_e/2$, and time scale is the inverse of the earth's rotation rate, $\Omega^{-1}$. The the $\beta-$ effect is zero in the isotropic integrations and $1.15\times10^{-11}\ \mathrm{m^{-1}s^{-1}}$ (corresponding to $\beta = \frac{1}{2}$ in non-dimensional units) in the anisotropic runs. This is characteristic of the earth's differential rotation at $60^{\circ}$ latitude, and the parameter $k_0^2 = \frac{1}{2}$ in Equation (6). We use a (bare) viscosity coefficient $\nu_0$ of $1.85\times10^{6}\ \mathrm{m^2s^{-1}}$ (non-dimensional $\nu_0 = 2.5\times10^{-3}$). In the closure integrations the mean eastward flow $U$ is either zero or has a speed of $15\ \mathrm{m\ s^{-1}}$ (non-dimensional $U = 0.065$). The closure runs are all unforced with $f_0 = 0$ in Equation (6) and the drag on the large-scale flow $\alpha_U = 0$ in Equation (5). The parameter $\gamma$ in Equation (72) that specifies the strength of the eddy damping is specified at $\gamma = 0.6$ as used in [9]. We have however checked that the broad conclusions regarding the similarity of the closure calculations described in Subsection 10.3 to follow hold for a wide range of $\gamma$ down to 0.01. The closure calculations are performed at a resolution of circular truncation C64 where $|\mathbf{k}| = k \le 64$. All integrations proceed to $t_{\max} = 90\ \mathrm{s}$ (non-dimensional $t_{\max} = 0.4$) with a time step of $\Delta t = 0.9$ s (non-dimensional $\Delta t = 0.004$). The time stepping is performed with a predictor-corrector scheme.

The closure calculations start from Gaussian initial conditions with $C_{\mathbf{k}}(t_0, t_0) = C_{\mathbf{k}}(0,0)$ for spectrum B of Frederiksen and Davies [15,16]:

$$C_k(0,0) = 0.18k^2 \exp\left(-\tfrac{2}{3}k\right). \tag{103}$$

From this spectrum the closure integrations proceed with one run being for isotropic turbulence denoted run ***I*** and three runs for anisotropic turbulence with Rossby waves denoted runs ***A1*** to ***A3***. The particular values of nondimensional $\beta$, $c$, of Equation (78), and $U$ that are used in runs ***I*** and ***A1*** to ***A3*** are given in Table 1.

**Table 1.** Non-dimensional parameters specifying the isotropic and three anisotropic closure runs.

| Closure Runs | $\beta$ | $c$ | $U$ |
|---|---|---|---|
| **Isotropic Run *I*** | 0 | 0 | 0 |
| **Anisotropic Run *A1*** | 0.5 | 0 | 0 |
| **Anisotropic Run *A2*** | 0.5 | 0.5 | 0 |
| **Anisotropic Run *A3*** | 0.5 | 0.5 | 0.065 |

*10.3 Moderate Reynolds Number Closure Integrations*

Next, we consider the evolution of the statistics of turbulence, possibly interacting with Rossby waves and a large-scale mean flow $U$ within the EDQNM and EDMAC models described in Section 7. For the isotropic run ***I***, and the anisotropic run ***A1***, $c = 0$, $U = 0$, and the EDQNM closure has been used for these. The EDQNM is not guaranteed to be realizable for the anisotropic run with $\beta = 0.5$. However, the EDQNM remained stable throughout the integration. Indeed, we found little difference in the results for anisotropic run ***A1*** with the EDQNM and anisotropic run ***A2*** with $c = 0.5$ performed with the EDMAC model, which is guaranteed to be realizable in the presence of Rossby waves. This is shown in Table 2 for the large-scale Reynolds number, and for the skewness, which are identical or essentially the same for runs ***I***, ***A1*** and ***A2***. The close similarity between these runs for the evolved transient kinetic energy and palinstrophy is also seen from Figures 1 and 2. In fact, the results are so close that we have had to shift down the plots by increasing factors of 10 in order separate the results.

The inclusion of the large-scale mean flow $U = 0.065$, corresponding to a dimensional value of $U = 15\ \mathrm{ms}^{-1}$, in the anisotropic run ***A3*** with the EDMAC model increases the evolved Reynolds number slightly and decrease the evolved skewness more. Thus, Rossby waves, in the presence of the large-scale flow, make the turbulence more Gaussian.

**Table 2.** Initial and evolved Reynolds number and skewness for the isotropic and three anisotropic closure runs.

| Closure Runs | $R_L(0)$ | $R_L(0.4)$ | $S(0)$ | $S(0.4)$ |
|---|---|---|---|---|
| **Isotropic Run *I*** | 304.8 | 263.7 | 0 | 0.735 |
| **Anisotropic Run *A1*** | 304.8 | 263.8 | 0 | 0.734 |
| **Anisotropic Run *A2*** | 304.8 | 263.8 | 0 | 0.735 |
| **Anisotropic Run *A3*** | 304.8 | 265.7 | 0 | 0.690 |

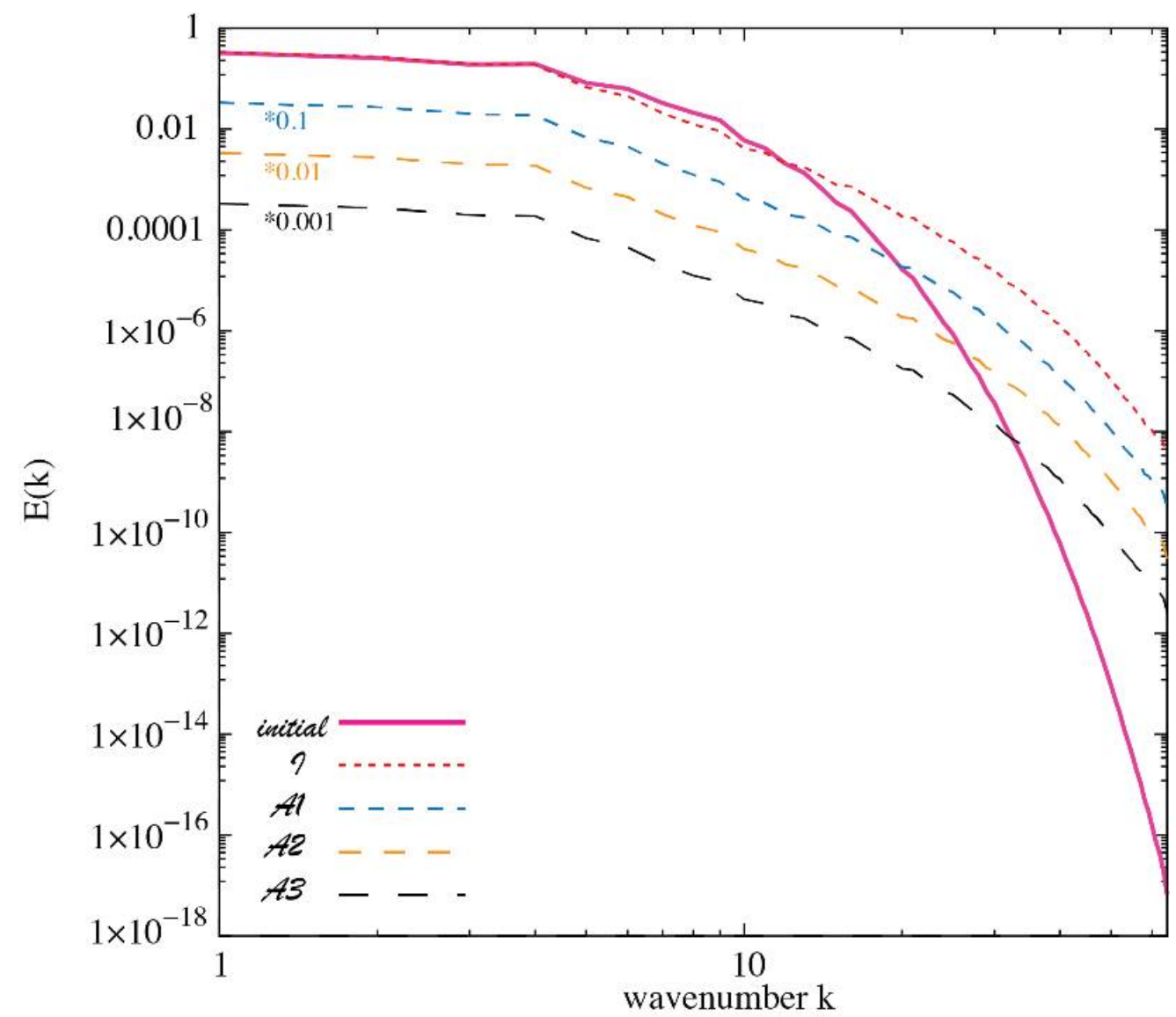

**Figure 1.** Comparison of transient kinetic energy spectra $E(k)$ for the EDQNM and EDMAC models for the runs in Table 2 at the initial ($t=0$) and final times ($t=0.4$).

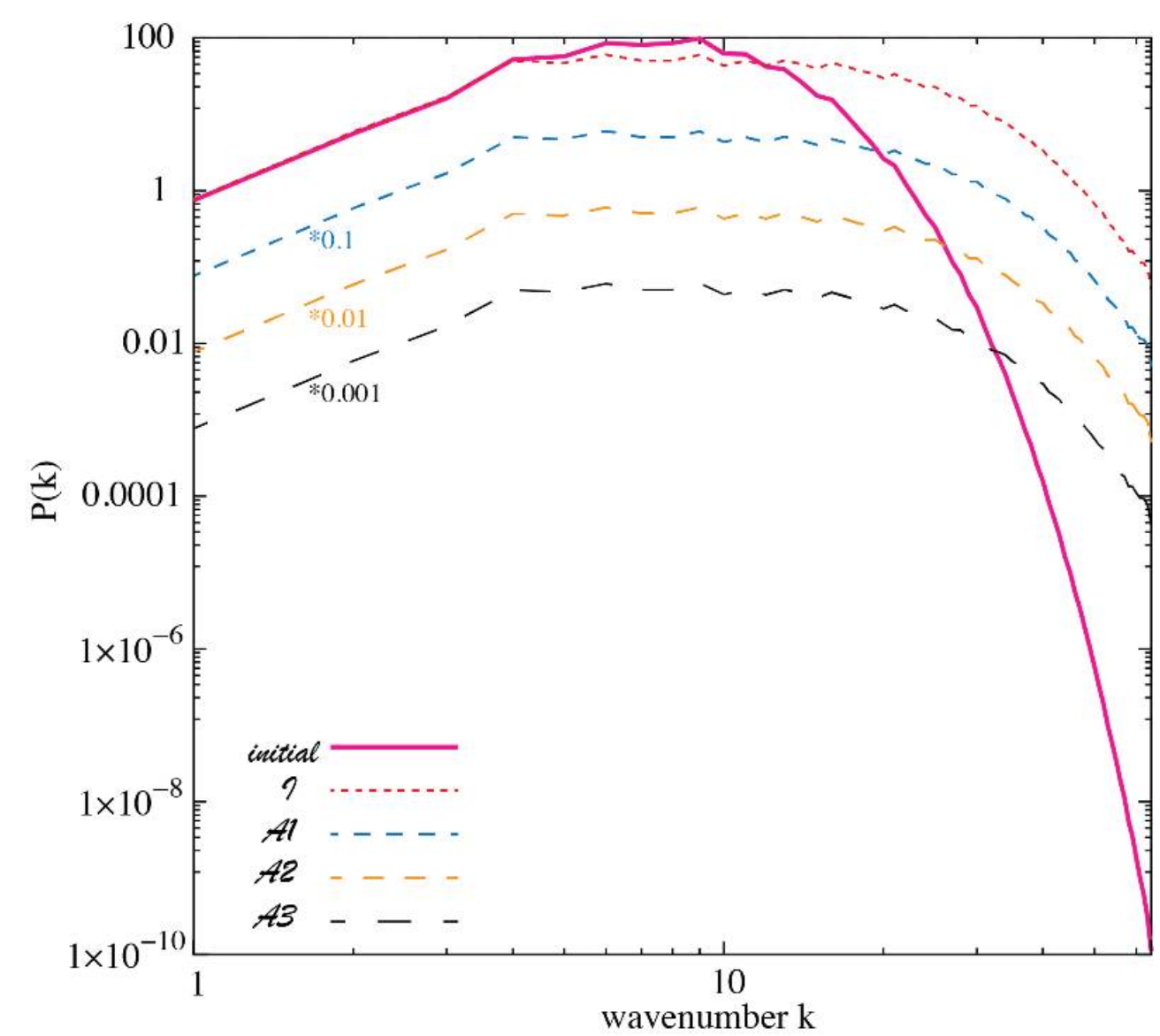

**Figure 2.** As in Figure 1 for the palinstrophy spectra $P(k)$.

Thus, in these closure integrations we conclude that the addition of the frequency-dependent damping in the EDMAC model does not greatly change the results compared with the EDQNM. This is the case for the strength of the eddy damping $\gamma = 0.6$ used in the displayed results. We have also checked that applies equally for strengths of $\gamma$ down to 0.01. As $\gamma$ is increased to larger values the frequency-dependent contribution to the eddy damping becomes smaller and the damping experienced by the EDMAC is closer to that of the EDQNM. While the EDQNM closure remained stable for the anisotropic run ***A1*** this cannot always be guaranteed for the EDQNM but it can for the EDMAC model under the conditions determined in Section 8 and Appendix B.

## 11. Discussion and Conclusions

The theoretical development of the Eddy Damped Markovian Anisotropic Closure (EDMAC) has been presented for anisotropic turbulence interacting with Rossby waves in the presence of advection by a large-scale mean flow. The EDMAC generalizes the Eddy Damped Quasi Normal Markovian (EDQNM) to a form that is realizable in the presence of transient waves. We have documented the equations for two-dimensional turbulence interacting with Rossby waves and a large-scale flow in physical space and in discrete Fourier spectral space.

### *11.1 Theoretical Results*

The development of the EDMAC model has then proceeded in a number of steps requiring the formulation of generalized non-Markovian and Markovian closures.

#### 11.1.1 Generalized non-Markovian Closures

Firstly, renormalized perturbation theory has been employed to construct a generalization of the non-Markovian Direct Interaction Approximation (DIA) closure with additional eddy damping and nonlinear noise terms that depend on the Doppler shifted Rossby wave frequencies. Associated generalized Self Consistent Field Theory (SCFT) and Local Energy-Transfer Theory (LET) closures have also been introduced. The next step has been to slightly simplify these non-Markovian closures to abridged forms in which the large-scale mean flow is regarded as slowly varying in the time history integrals.

#### 11.1.2 Markovian Anisotropic Closures

Markovian Anisotropic Closures (MACs) have then been developed by employing any of three forms of the Fluctuation Dissipation Theorem in Equation (4). This simplifies the time history integrals to the extent that their effects can be encapsulated in two relaxation functions that can also be determined by ordinary Markovian differential equations. While the MACs are more efficient for long time integrations than the non-Markovian closures, the calculation of the relaxation functions through differential equations is a considerable overhead compared with the analytical forms that are used for the EDQNM and established here for the EDMAC.

#### 11.1.3 Realizable Eddy Damped Markovian Anisotropic Closure

The EDMAC has been derived from the MAC employing the current-time Fluctuation Dissipation Theorem (FDT) in Equation (1). Further simplification of the relaxation functions has been achieved by making the Markov approximation that the generalized dissipation functionals are slowly varying in the time history integrals, as for the EDQNM [1]. The consequent EDMAC model has the same eddy damping term as in the EDQNM that parameterizes turbulent mixing [1,2,50,51], but also has a frequency-dependent damping which ensures the realizability of the EDMAC in the presence of transient waves. The strength of the frequency dependent damping has been determined under the assumption of small amplitude waves, again based on renormalized perturbation theory. More generally, the strength may be specified empirically and sufficient conditions on the frequency-dependent damping that ensure realizability of the EDMAC have been determined.

#### 11.1.4 Generalizations of the Eddy Damped Markovian Anisotropic Closure

We have considered the relationships between EDQNM and EDMAC models for anisotropic turbulence interacting with transient waves for systems involving higher order nonlinearity than quadratic and in higher dimension than two. The conditions that ensure that the real part of the EDMAC relaxation function is again positive semi-definite have been determined.

### *11.2 Numerical Closure Calculations*

In this, largely theoretical, paper we have also reported in detail on four integrations of the closures for rapidly evolving turbulence. They have been performed for isotropic 2D turbulence, for anisotropic turbulence with Rossby waves with and without the frequency-dependent damping and for anisotropic turbulence with Rossby waves and large-scale mean flow and frequency-dependent damping. In these numerical experiments, with empirical parameter $\gamma = 0.6$ in Equation (72), the results of the evolved simulations with EDQNM and EDMAC turn out very similar to each other with little effect on one-dimensional spectra. This is also so for smaller $\gamma$ down to $\gamma = 0.01$ while for larger $\gamma$ the frequency-dependent drain becomes smaller and the total drain in the EDMAC approaches that of the EDQNM.

*11.3 Conclusions and Future Prospects*

The realizability of the EDMAC in the presence of waves at little or no extra computational cost over the EDQNM should make it attractive provided the additional frequency-dependent damping does not change the broad properties of the numerical integrations with the closures. We have noted in Section 9, the prospects for generalizing the realizable EDMAC model to higher dimensions and different physical systems including with higher order nonlinearity. In future studies we plan to explore the performance of the EDMAC model in various settings. As well we aim to formulate and study the performance of a similar inhomogeneous closure, the Eddy Damped Markovian Inhomogeneous Closure (EDMIC) [45], generalized to be realizable for turbulence interacting with transient Rossby waves.

**Author Contributions:** Conceptualization, JSF and TJO; methodology, JSF; formal analysis, JSF; numeric implementation, TJO and JSF; investigation, JSF and TJO; writing—original draft preparation, JSF; writing—review and editing, JSF and TJO; supervision, JSF; project administration, JSF; funding acquisition, TJO. All authors have read and agreed to the published version of the manuscript.

**Funding:** TJO was funded by CSIRO Environment.

**Data Availability Statement:** All data displayed in figures 1 & 2 is available on request.

**Conflicts of Interest:** The authors declare no conflict of interest.

## Appendix A: Renormalized Perturbation Theory

The aim of this Appendix is to provide motivation for the modified form of the eddy damping used for the realizable EDMAC model. We begin by expanding $\tilde{\zeta}_{\mathbf{k}}$ in a perturbation series where, from Equations (11), (16), and (17),

$$\begin{aligned}&\left(\frac{\partial}{\partial t}+\nu_0(k)k^2+\mathrm{i}\lambda\omega_{\mathbf{k}}^U(t)\right)\tilde{\zeta}_{\mathbf{k}}(t)\\&=\lambda\sum_{\mathbf{p}}\sum_{\mathbf{q}}\delta(\mathbf{k},\mathbf{p},\mathbf{q})K(\mathbf{k},\mathbf{p},\mathbf{q})\tilde{\zeta}_{-\mathbf{p}}(t)\tilde{\zeta}_{-\mathbf{q}}(t)+\tilde{f}_0(\mathbf{k},t).\end{aligned} \tag{A1}$$

That is,

$$\tilde{\zeta}_{\mathbf{k}} = \tilde{\zeta}_{\mathbf{k}}^{(0)} + \lambda\tilde{\zeta}_{\mathbf{k}}^{(1)} + \ldots \tag{A2}$$

where in both Equations (A1) and (A2) we have introduced the book keeping small parameter $\lambda$ that will be set to unity after the renormalization process.

In Equation (A1) $\omega_{\mathbf{k}}^U$ is the Doppler shifted Rossby wave frequency defined in Equation (13). We note that $\omega_{\mathbf{k}}^U$ can also be written in the form:

$$\begin{aligned}\mathrm{i}\omega_{\mathbf{k}}^U(t)&=\mathrm{i}k_0U(t)\frac{k_0k_x(k^2-k_0^2)}{k^2k_0^2}-\mathrm{i}k_0^{-1}\beta\frac{k_0k_x}{k^2}\\&=-[2K(\mathbf{k},-\mathbf{k},\mathbf{0})\zeta_{-\mathbf{0}}+A(\mathbf{k},-\mathbf{k},\mathbf{0})h_{-\mathbf{0}}^{\beta}]\end{aligned} \tag{A3}$$

where the interaction coefficients are defined by

$$A(\mathbf{k},-\mathbf{k},\mathbf{0}) = k_0k_x / k^2 \tag{A4}$$

and

$$K(\mathbf{k},-\mathbf{k},\mathbf{0})=\tfrac{1}{2}[A(\mathbf{k},-\mathbf{k},\mathbf{0})+A(\mathbf{k},\mathbf{0},-\mathbf{k})]=-\tfrac{1}{2}k_0k_x(k^2-k_0^2)/k^2k_0^2. \quad \text{(A5)}$$

As well we have defined

$$\zeta_{-\mathbf{0}}=\mathrm{i}k_0U,\ \zeta_{\mathbf{0}}=\zeta_{-\mathbf{0}}^{*} \quad \text{(A6)}$$

and

$$h_{-\mathbf{0}}^{\beta}=\mathrm{i}k_0^{-1}\beta,\ h_{\mathbf{0}}^{\beta}=(h_{-\mathbf{0}}^{\beta})^{*}. \quad \text{(A7)}$$

Here we have been motivated by the study of Frederiksen and O'Kane [47]. There the transformation in Equation (A6), and the suitable definitions of the interaction coefficients in Equations (A4) and (A5), allowed the combination of the equations for the small scales and the large scale flow into a single form. Here, our aim has been to demonstrate that the term involving the Doppler shifted Rossby wave frequency in Equation (A1) is proportional to the interaction coefficients and, as for the other terms multiplying interaction coefficients there, is assumed to be order $\lambda$ in the perturbation theory.

From Equations (A1) and (A2) we see that to zero order in perturbation theory we have

$$\left(\frac{\partial}{\partial t}+\nu_0(k)k^2\right)\tilde{\zeta}_{\mathbf{k}}^{(0)}(t)=\tilde{f}_0(\mathbf{k},t). \quad \text{(A8)}$$

To first order we have

$$\begin{aligned}\left(\frac{\partial}{\partial t}+\nu_0(k)k^2\right)\tilde{\zeta}_{\mathbf{k}}^{(1)}(t)&=-\mathrm{i}\omega_{\mathbf{k}}^{U}(t)\tilde{\zeta}_{\mathbf{k}}^{(0)}(t)\\ &+\sum_{\mathbf{p}}\sum_{\mathbf{q}}\delta(\mathbf{k},\mathbf{p},\mathbf{q})K(\mathbf{k},\mathbf{p},\mathbf{q})\tilde{\zeta}_{-\mathbf{p}}^{(0)}(t)\tilde{\zeta}_{-\mathbf{q}}^{(0)}(t).\end{aligned} \quad \text{(A9)}$$

Then, the formal solution to Equation (A9) can be written, using the Greens function $R_{\mathbf{k}}^{(0)}(t,s)\equiv R_{\mathbf{k},\mathbf{k}}^{(0)}(t,s)$ corresponding to Equation (A8), as follows

$$\begin{aligned}\tilde{\zeta}_{\mathbf{k}}^{(1)}(t)=\int_{t_0}^{t}ds\,R_{\mathbf{k}}^{(0)}(t,s)\Big\{&-\mathrm{i}\omega_{\mathbf{k}}^{U}(s)\tilde{\zeta}_{\mathbf{k}}^{(0)}(s)\\ &+\sum_{\mathbf{p}}\sum_{\mathbf{q}}\delta(\mathbf{k},\mathbf{p},\mathbf{q})K(\mathbf{k},\mathbf{p},\mathbf{q})\tilde{\zeta}_{-\mathbf{p}}^{(0)}(s)\tilde{\zeta}_{-\mathbf{q}}^{(0)}(s)\Big\}.\end{aligned} \quad \text{(A10)}$$

The two-time cumulant can also be expressed in a perturbation series. Here, we consider homogenous turbulence and

$$\begin{aligned}&C_{\mathbf{k},-\mathbf{k}}(t,t')\equiv C_{\mathbf{k}}(t,t')=<\tilde{\zeta}_{\mathbf{k}}(t)\tilde{\zeta}_{-\mathbf{k}}(t')>\\ &=<\tilde{\zeta}_{\mathbf{k}}^{(0)}(t)\tilde{\zeta}_{-\mathbf{k}}^{(0)}(t')>+\lambda<\tilde{\zeta}_{\mathbf{k}}^{(1)}(t)\tilde{\zeta}_{-\mathbf{k}}^{(0)}(t')>+\lambda<\tilde{\zeta}_{\mathbf{k}}^{(0)}(t)\tilde{\zeta}_{-\mathbf{k}}^{(1)}(t')>+\ldots\end{aligned} \quad \text{(A11)}$$

To first order in $\lambda$ we have the two-time cumulant contribution

$$\begin{aligned}C_{\mathbf{k}}^{(1)}(t,t')&=\int_{t_0}^{t}ds\,R_{\mathbf{k}}^{(0)}(t,s)<\tilde{\zeta}_{\mathbf{k}}^{(0)}(s)\hat{\zeta}_{-\mathbf{k}}^{(0)}(t')>[-\mathrm{i}\omega_{\mathbf{k}}^{U}(s)]\\ &+\int_{t_0}^{t'}ds\,R_{-\mathbf{k}}^{(0)}(t',s)<\tilde{\zeta}_{-\mathbf{k}}^{(0)}(s)\hat{\zeta}_{\mathbf{k}}^{(0)}(t)>[-\mathrm{i}\omega_{-\mathbf{k}}^{U}(s)]\\ &=\int_{t_0}^{t}ds\,R_{\mathbf{k}}^{(0)}(t,s)C_{-\mathbf{k}}^{(0)}(t',s)[-\mathrm{i}\omega_{\mathbf{k}}^{U}(s)]+\int_{t_o}^{t'}ds\,R_{-\mathbf{k}}^{(0)}(t',s)C_{\mathbf{k}}^{(0)}(t,s)[-\mathrm{i}\omega_{-\mathbf{k}}^{U}(s)].\end{aligned} \quad \text{(A12)}$$

Next, we consider the perturbation expansion for the response function

$$R_{\mathbf{k},\mathbf{k}}(t,t') \equiv R_{\mathbf{k}}(t,t') = \left\langle \frac{\delta\tilde{\zeta}_{\mathbf{k}}^{(0)}(t)}{\delta\tilde{f}_{\mathbf{k}}(t')} \right\rangle + \lambda \left\langle \frac{\delta\tilde{\zeta}_{\mathbf{k}}^{(1)}(t)}{\delta\tilde{f}_{\mathbf{k}}(t')} \right\rangle + \ldots$$
$$= \left\langle \tilde{R}_{\mathbf{k}}^{(0)}(t,t') \right\rangle + \lambda \left\langle \tilde{R}_{\mathbf{k}}^{(1)}(t,t') \right\rangle + \ldots \quad \text{(A13)}$$

where $\tilde{R}_{\mathbf{k}}^{(0)}(t,t')$ is the response function corresponding to Equation (A8) and $R_{\mathbf{k}}^{(0)}(t,t') = \left\langle \tilde{R}_{\mathbf{k}}^{(0)}(t,t') \right\rangle$ is the corresponding average. To first order in $\lambda$, we have

$$R_{\mathbf{k}}^{(1)}(t,t') = \int_{t'}^{t} ds\, R_{\mathbf{k}}^{(0)}(t,s) R_{\mathbf{k}}^{(0)}(s,t')[-\mathrm{i}\omega_{\mathbf{k}}^{U}(s)]. \quad \text{(A14)}$$

We are now able to write down the dynamical equation for the response function to second order in the perturbation parameter $\lambda$:

$$\begin{aligned}
&\left(\frac{\partial}{\partial t} + \nu_0(k)k^2\right) R_{\mathbf{k}}(t,t') - \delta(t-t') \\
&= \lambda[R_{\mathbf{k}}^{(0)}(t,t') + \lambda R_{\mathbf{k}}^{(1)}(t,t')][-\mathrm{i}\omega_{\mathbf{k}}^{U}(t)] \\
&+2\lambda^2 \sum_{\mathbf{p}} \sum_{\mathbf{q}} \delta(\mathbf{k},\mathbf{p},\mathbf{q}) K(\mathbf{k},\mathbf{p},\mathbf{q}) \\
&\quad \times \left\{ < \tilde{R}_{-\mathbf{p}}^{(0)}(t,t') \zeta_{-\mathbf{q}}^{(1)}(t) > + < \tilde{R}_{-\mathbf{p}}^{(1)}(t,t') \zeta_{-\mathbf{q}}^{(0)}(t) > \right\} \\
&= \lambda[-\mathrm{i}\omega_{\mathbf{k}}^{U}(t)] R_{\mathbf{k}}^{(0)}(t,t') + \lambda^2[-\mathrm{i}\omega_{\mathbf{k}}^{U}(t)] \int_{t'}^{t} ds\, R_{\mathbf{k}}^{(0)}(t,s)[-\mathrm{i}\omega_{\mathbf{k}}^{U}(s)] R_{\mathbf{k}}^{(0)}(s,t')] \\
&+4\lambda^2 \int_{t'}^{t} ds \sum_{\mathbf{p}} \sum_{\mathbf{q}} \delta(\mathbf{k},\mathbf{p},\mathbf{q}) K(\mathbf{k},\mathbf{p},\mathbf{q}) K(-\mathbf{p},-\mathbf{q},-\mathbf{k}) \\
&\quad \times R_{-\mathbf{p}}^{(0)}(t,s) C_{-\mathbf{q}}^{(0)}(t,s) R_{-\mathbf{k}}^{(0)}(s,t')
\end{aligned} \quad \text{(A15)}$$

with $R_{\mathbf{k}}(t,t) = 1$. Perturbation expansion of the last term in Equation (A15) is as described in Frederiksen [9]. On renormalizing with $\lambda \to 1$ and the zero-order terms replaced by the renormalized terms we have

$$\begin{aligned}
&\left(\frac{\partial}{\partial t} + \nu_0(k)k^2 + \mathrm{i}\omega_{\mathbf{k}}^{U}(t)\right) R_{\mathbf{k}}(t,t') \\
&+ \int_{t'}^{t} ds(\eta_{\mathbf{k}}(t,s) + \pi_{\mathbf{k}}^{\omega}(t,s)) R_{\mathbf{k}}(s,t') = \delta(t-t')
\end{aligned} \quad \text{(A16)}$$

where the delta function ensures the initial condition $R_{\mathbf{k}}(t,t) = 1$ and

$$\pi_{\mathbf{k}}^{\omega}(t,s) = R_{\mathbf{k}}(t,s)\omega_{\mathbf{k}}^{U}(t)\omega_{\mathbf{k}}^{U}(s) \quad \text{(A17)}$$

as also shown in Equation (26). Here the nonlinear damping $\eta_{\mathbf{k}}(t,s)$ is given by

$$\eta_{\mathbf{k}}(t,s) = -4\sum_{\mathbf{p}} \sum_{\mathbf{q}} \delta(\mathbf{k},\mathbf{p},\mathbf{q}) K(\mathbf{k},\mathbf{p},\mathbf{q}) K(-\mathbf{p},-\mathbf{q},-\mathbf{k}) R_{-\mathbf{p}}(t,s) C_{-\mathbf{q}}(t,s), \quad \text{(A18)}$$

as also shown in in Equation (24).

Next, we consider the two-time two-point cumulant equation which takes the form:

$$
\begin{aligned}
&\left(\frac{\partial}{\partial t}+\nu_0(k)k^2\right)C_{\mathbf{k}}(t,t') \\
&=\lambda[C_{\mathbf{k}}^{(0)}(t,t')+\lambda C_{\mathbf{k}}^{(1)}(t,t')][-\mathrm{i}\omega_{\mathbf{k}}^U(t)]+<\tilde{f}_0(\mathbf{k},t)\tilde{\zeta}_{-\mathbf{k}}(t')> \\
&+\lambda\sum_{\mathbf{p}}\sum_{\mathbf{q}}\delta(\mathbf{k},\mathbf{p},\mathbf{q})K(\mathbf{k},\mathbf{p},\mathbf{q}) \\
&\quad\times\left\{2\lambda<\tilde{\zeta}_{-\mathbf{p}}^{(1)}(t)\,\tilde{\zeta}_{-\mathbf{q}}^{(0)}(t)\tilde{\zeta}_{-\mathbf{k}}^{(0)}(t')>+\lambda<\tilde{\zeta}_{-\mathbf{p}}^{(0)}(t)\,\tilde{\zeta}_{-\mathbf{q}}^{(0)}(t)\tilde{\zeta}_{-\mathbf{k}}^{(1)}(t')>\right\} \\
&=\lambda[-\mathrm{i}\omega_{\mathbf{k}}^U(t)]C_{\mathbf{k}}^{(0)}(t,t')+\lambda^2[-\mathrm{i}\omega_{\mathbf{k}}^U(t)]\int_{t_0}^{t}ds\,R_{\mathbf{k}}^{(0)}(t,s)C_{-\mathbf{k}}^{(0)}(t',s)[-\mathrm{i}\omega_{\mathbf{k}}^U(s)] \\
&+\lambda^2[-\mathrm{i}\omega_{\mathbf{k}}^U(t)]\int_{t_0}^{t'}ds\,R_{-\mathbf{k}}^{(0)}(t',s)C_{\mathbf{k}}^{(0)}(t,s)[-\mathrm{i}\omega_{-\mathbf{k}}^U(s)]+<\tilde{f}_0(\mathbf{k},t)\tilde{\zeta}_{-\mathbf{k}}(t')> \\
&+4\lambda^2\int_{t_0}^{t}ds\sum_{\mathbf{p}}\sum_{\mathbf{q}}\delta(\mathbf{k},\mathbf{p},\mathbf{q})K(\mathbf{k},\mathbf{p},\mathbf{q})K(-\mathbf{p},-\mathbf{q},-\mathbf{k}) \\
&\quad\times R_{-\mathbf{p}}^{(0)}(t,s)C_{-\mathbf{q}}^{(0)}(t,s)C_{-\mathbf{k}}^{(0)}(t',s) \\
&+2\lambda^2\int_{t_0}^{t'}ds\sum_{\mathbf{p}}\sum_{\mathbf{q}}\delta(\mathbf{k},\mathbf{p},\mathbf{q})K(\mathbf{k},\mathbf{p},\mathbf{q})K(-\mathbf{k},-\mathbf{p},-\mathbf{q}) \\
&\quad\times C_{-\mathbf{p}}^{(0)}(t,s)C_{-\mathbf{q}}^{(0)}(t,s)R_{-\mathbf{k}}^{(0)}(t',s).
\end{aligned}
\tag{A19}
$$

Again, the perturbation expansion of the three-point terms in Equation (A19) is as described in Frederiksen [9]. Finally, renormalizing with $\lambda\to 1$ and the zero-order terms replaced by the renormalized terms we have

$$
\begin{aligned}
&\left(\frac{\partial}{\partial t}+\nu_0(k)k^2+\mathrm{i}\omega_{\mathbf{k}}^U(t)\right)C_{\mathbf{k}}(t,t') \\
&+\int_{t_0}^{t}ds\,(\eta_{\mathbf{k}}(t,s)+\pi_{\mathbf{k}}^{\omega}(t,s))C_{-\mathbf{k}}(t',s) \\
&=\int_{t_0}^{t'}ds\,(S_{\mathbf{k}}(t,s)+P_{\mathbf{k}}^{\omega}(t,s)+F_0(\mathbf{k},t,s))R_{-\mathbf{k}}(t',s).
\end{aligned}
\tag{A20}
$$

Here, the bare noise $F_0(\mathbf{k},t,s)$ is given in Equation (23) and

$$P_{\mathbf{k}}^{\omega}(t,s)=C_{\mathbf{k}}(t,s)\omega_{\mathbf{k}}^U(t)\omega_{\mathbf{k}}^U(s) \tag{A21}$$

as also shown in Equation (27) with $\pi_{\mathbf{k}}^{\omega}(t,s)$ given in Equation (A17) and $\eta_{\mathbf{k}}(t,s)$ in Equation (A18). The nonlinear noise $S_{\mathbf{k}}(t,s)$ takes the form:

$$S_{\mathbf{k}}(t,s)=2\sum_{\mathbf{p}}\sum_{\mathbf{q}}\delta(\mathbf{k},\mathbf{p},\mathbf{q})K(\mathbf{k},\mathbf{p},\mathbf{q})K(-\mathbf{k},-\mathbf{p},-\mathbf{q})C_{-\mathbf{p}}(t,s)C_{-\mathbf{q}}(t,s) \tag{A22}$$

as also given in Equation (25).

To close the system of equations we also need the single-time cumulant equation which takes the form:

$$\left(\frac{\partial}{\partial t}+2\nu_0(k)k^2\right)C_{\mathbf{k}}(t,t)$$
$$+2\,\mathrm{Re}\int_{t_0}^{t} ds\,(\eta_{\mathbf{k}}(t,s)+\pi_{\mathbf{k}}^{\omega}(t,s))C_{-\mathbf{k}}(t,s) \tag{A23}$$
$$=2\,\mathrm{Re}\int_{t_0}^{t} ds\,(S_{\mathbf{k}}(t,s)+P_{\mathbf{k}}^{\omega}(t,s)+F_0(\mathbf{k},t,s))R_{-\mathbf{k}}(t,s).$$

Here, the single-time cumulant $C_{\mathbf{k}}(t_0,t_0)$ is to be specified as the initial condition.

**Appendix B: Langevin Equation Underpinning the EDMAC Model**

In this Appendix we consider the realizability of the EDMAC model for the single-time cumulant in Equation (61), with $\Theta^{EDMAC}$ replacing $\Theta^{X=0}$. The approach follows that of Leith [3] and Herring and Kraichnan [60] in establishing the realizability of the EDQNM. Again, EDMAC can be shown to be realizable by being underpinned by a suitable stochastic model. The EDMAC model can be constructed exactly from the following Langevin equation:

$$\left(\frac{\partial}{\partial t}+D_0(\mathbf{k},t)+D_\eta(\mathbf{k},t)\right)\tilde{\zeta}_{\mathbf{k}}(t)=f_0(\mathbf{k},t)+f_S(\mathbf{k},t) \tag{A24}$$

Here,

$$D_0(\mathbf{k},t)=\nu_0(k)k^2+\mathrm{i}\omega_{\mathbf{k}}^{U}(t) \tag{A25}$$

as in Equation (44) but with

$$D_\eta(\mathbf{k},t)$$
$$=-4\sum_{\mathbf{p}}\sum_{\mathbf{q}}\delta(\mathbf{k},\mathbf{p},\mathbf{q})K(\mathbf{k},\mathbf{p},\mathbf{q})K(\mathbf{p},\mathbf{q},\mathbf{k})C_{\mathbf{q}}(t,t)\Theta^{EDMAC}(\mathbf{k},\mathbf{p},\mathbf{q})(t), \tag{A26}$$

also displayed in Equation (64), and with $\Theta^{EDMAC}$ replacing $\Theta^{X=0}$. As shown in Equation (83),

$$\Theta^{EDMAC}(\mathbf{k},\mathbf{p},\mathbf{q})(t)$$
$$=\frac{1-\exp\left(-\left[\rho_{\mathbf{k}}(t)+\rho_{\mathbf{p}}(t)+\rho_{\mathbf{q}}(t)+\mathrm{i}(\omega_{\mathbf{k}}^{U}(t)+\omega_{\mathbf{p}}^{U}(t)+\omega_{\mathbf{q}}^{U}(t))\right](t-t_0)\right)}{\rho_{\mathbf{k}}(t)+\rho_{\mathbf{p}}(t)+\rho_{\mathbf{q}}(t)+\mathrm{i}(\omega_{\mathbf{k}}^{U}(t)+\omega_{\mathbf{p}}^{U}(t)+\omega_{\mathbf{q}}^{U}(t))}. \tag{A27}$$

In Equation (A24),

$$f_0(\mathbf{k},t)=\tilde{f}_0(\mathbf{k},t), \tag{A28}$$

the bare random forcing and

$$f_S(\mathbf{k},t)=\sqrt{2}\sum_{\mathbf{p}}\sum_{\mathbf{q}}\delta(\mathbf{k}+\mathbf{p}+\mathbf{q})K(\mathbf{k},\mathbf{p},\mathbf{q})\left[\mathrm{Re}\,\Theta^{EDMAC}(\mathbf{k},\mathbf{p},\mathbf{q})(t)\right]^{\frac{1}{2}}$$
$$\times w(t)W_{-\mathbf{p}}^{(1)}(t)W_{-\mathbf{q}}^{(2)}(t). \tag{A29}$$

which determines the nonlinear noise. Equation (A29) contains the independent random variables $W_{\mathbf{k}}^{(i)}(t)$, where $i$=1, 2 or 3, and $w(t)$. Their properties are such that the statistical dynamical closure for the single-time cumulant in Equation (61), but with $\Theta^{EDMAC}$ replacing $\Theta^{X=0}$, is reconstructed. The statistics of the random variables are such that:

$$<W_{\mathbf{k}}^{(i)}(t)W_{-\mathbf{l}}^{(j)}(t')>=\delta_{ij}\delta_{\mathbf{kl}}C_{\mathbf{k}}(t,t'), \tag{A30}$$

with

$$<\tilde{\zeta}_{\mathbf{k}}(t)\tilde{\zeta}_{-\mathbf{k}}(t')>=C_{\mathbf{k}}(t,t'), \tag{A31}$$

and

$$< w(t)w(t') >= \delta(t-t'). \tag{A32}$$

Here, $\delta_{ij}$ and $\delta_{\mathbf{kl}}$ are Kronecker delta functions and $\delta(t-t')$ is the Dirac delta function.

The single-time cumulants $C_{\mathbf{k}}(t,t)$, in the EDMAC model, that are determined by Equation (61), but with $\Theta^{EDMAC}$ replacing $\Theta^{X=0}$, are realizable. This follows from the Langevin Equation (A24) provided $\mathrm{Re}\,\Theta^{EDMAC}(\mathbf{k},\mathbf{p},\mathbf{q})(t) \geq 0$. Section 8 outlines the demonstration of the positivity of $\mathrm{Re}\,\Theta^{EDMAC}$ provided $c \geq \frac{1}{4}$. In the inviscid unforced case, the EDMAC equations conserve kinetic energy and enstrophy.